%Paper: cond-mat/9408092
%From: fabman@avzw03.physik.uni-bonn.de (Fabian Essler)
%Date: Mon, 29 Aug 1994 22:29:08 +0200

% File jytex.tex, for jyTeX version 2.6M (June 1992)
% Copyright (c) 1991, 1992 by Jonathan P. Yamron

\catcode`\@=11

%******************************************************************************

\message{Loading jyTeX fonts...}

%************************************************************
%*
%*             Available fonts
%*
%************************************************************

%************** 5-point fonts *******************************

\font\vptrm=cmr5
\font\vptmit=cmmi5
\font\vptsy=cmsy5
\font\vptbf=cmbx5

\skewchar\vptmit='177 \skewchar\vptsy='60
\fontdimen16 \vptsy=\the\fontdimen17 \vptsy

\def\vpt{\ifmmode\err@badsizechange\else
     \@mathfontinit
     \textfont0=\vptrm  \scriptfont0=\vptrm  \scriptscriptfont0=\vptrm
     \textfont1=\vptmit \scriptfont1=\vptmit \scriptscriptfont1=\vptmit
     \textfont2=\vptsy  \scriptfont2=\vptsy  \scriptscriptfont2=\vptsy
     \textfont3=\xptex  \scriptfont3=\xptex  \scriptscriptfont3=\xptex
     \textfont\bffam=\vptbf
     \scriptfont\bffam=\vptbf
     \scriptscriptfont\bffam=\vptbf
     \@fontstyleinit
     \def\rm{\vptrm\fam=\z@}%
     \def\bf{\vptbf\fam=\bffam}%
     \def\oldstyle{\vptmit\fam=\@ne}%
     \rm\fi}

%************** 6-point fonts *******************************

\font\viptrm=cmr6
\font\viptmit=cmmi6
\font\viptsy=cmsy6
\font\viptbf=cmbx6

\skewchar\viptmit='177 \skewchar\viptsy='60
\fontdimen16 \viptsy=\the\fontdimen17 \viptsy

\def\vipt{\ifmmode\err@badsizechange\else
     \@mathfontinit
     \textfont0=\viptrm  \scriptfont0=\vptrm  \scriptscriptfont0=\vptrm
     \textfont1=\viptmit \scriptfont1=\vptmit \scriptscriptfont1=\vptmit
     \textfont2=\viptsy  \scriptfont2=\vptsy  \scriptscriptfont2=\vptsy
     \textfont3=\xptex   \scriptfont3=\xptex  \scriptscriptfont3=\xptex
     \textfont\bffam=\viptbf
     \scriptfont\bffam=\vptbf
     \scriptscriptfont\bffam=\vptbf
     \@fontstyleinit
     \def\rm{\viptrm\fam=\z@}%
     \def\bf{\viptbf\fam=\bffam}%
     \def\oldstyle{\viptmit\fam=\@ne}%
     \rm\fi}

%************** 7-point fonts *******************************

\font\viiptrm=cmr7
\font\viiptmit=cmmi7
\font\viiptsy=cmsy7
\font\viiptit=cmti7
\font\viiptbf=cmbx7

\skewchar\viiptmit='177 \skewchar\viiptsy='60
\fontdimen16 \viiptsy=\the\fontdimen17 \viiptsy

\def\viipt{\ifmmode\err@badsizechange\else
     \@mathfontinit
     \textfont0=\viiptrm  \scriptfont0=\vptrm  \scriptscriptfont0=\vptrm
     \textfont1=\viiptmit \scriptfont1=\vptmit \scriptscriptfont1=\vptmit
     \textfont2=\viiptsy  \scriptfont2=\vptsy  \scriptscriptfont2=\vptsy
     \textfont3=\xptex    \scriptfont3=\xptex  \scriptscriptfont3=\xptex
     \textfont\itfam=\viiptit
     \scriptfont\itfam=\viiptit
     \scriptscriptfont\itfam=\viiptit
     \textfont\bffam=\viiptbf
     \scriptfont\bffam=\vptbf
     \scriptscriptfont\bffam=\vptbf
     \@fontstyleinit
     \def\rm{\viiptrm\fam=\z@}%
     \def\it{\viiptit\fam=\itfam}%
     \def\bf{\viiptbf\fam=\bffam}%
     \def\oldstyle{\viiptmit\fam=\@ne}%
     \rm\fi}

%************** 8-point fonts *******************************

\font\viiiptrm=cmr8
\font\viiiptmit=cmmi8
\font\viiiptsy=cmsy8
\font\viiiptit=cmti8
%\font\viiiptsl=cmsl8
\font\viiiptbf=cmbx8
%\font\viiipttt=cmtt8
%\font\viiiptss=cmss8

\skewchar\viiiptmit='177 \skewchar\viiiptsy='60
\fontdimen16 \viiiptsy=\the\fontdimen17 \viiiptsy

\def\viiipt{\ifmmode\err@badsizechange\else
     \@mathfontinit
     \textfont0=\viiiptrm  \scriptfont0=\viptrm  \scriptscriptfont0=\vptrm
     \textfont1=\viiiptmit \scriptfont1=\viptmit \scriptscriptfont1=\vptmit
     \textfont2=\viiiptsy  \scriptfont2=\viptsy  \scriptscriptfont2=\vptsy
     \textfont3=\xptex     \scriptfont3=\xptex   \scriptscriptfont3=\xptex
     \textfont\itfam=\viiiptit
     \scriptfont\itfam=\viiptit
     \scriptscriptfont\itfam=\viiptit
     \textfont\bffam=\viiiptbf
     \scriptfont\bffam=\viptbf
     \scriptscriptfont\bffam=\vptbf
     \@fontstyleinit
     \def\rm{\viiiptrm\fam=\z@}%
     \def\it{\viiiptit\fam=\itfam}%
     \def\bf{\viiiptbf\fam=\bffam}%
     \def\oldstyle{\viiiptmit\fam=\@ne}%
     \rm\fi}

%************** Optional 9-point fonts **********************

\def\getixpt{%
     \font\ixptrm=cmr9
     \font\ixptmit=cmmi9
     \font\ixptsy=cmsy9
     \font\ixptit=cmti9
%     \font\ixptsl=cmsl9
     \font\ixptbf=cmbx9
%     \font\ixpttt=cmtt9
%     \font\ixptss=cmss9
     \skewchar\ixptmit='177 \skewchar\ixptsy='60
     \fontdimen16 \ixptsy=\the\fontdimen17 \ixptsy}

\def\ixpt{\ifmmode\err@badsizechange\else
     \@mathfontinit
     \textfont0=\ixptrm  \scriptfont0=\viiptrm  \scriptscriptfont0=\vptrm
     \textfont1=\ixptmit \scriptfont1=\viiptmit \scriptscriptfont1=\vptmit
     \textfont2=\ixptsy  \scriptfont2=\viiptsy  \scriptscriptfont2=\vptsy
     \textfont3=\xptex   \scriptfont3=\xptex    \scriptscriptfont3=\xptex
     \textfont\itfam=\ixptit
     \scriptfont\itfam=\viiptit
     \scriptscriptfont\itfam=\viiptit
     \textfont\bffam=\ixptbf
     \scriptfont\bffam=\viiptbf
     \scriptscriptfont\bffam=\vptbf
     \@fontstyleinit
     \def\rm{\ixptrm\fam=\z@}%
     \def\it{\ixptit\fam=\itfam}%
     \def\bf{\ixptbf\fam=\bffam}%
     \def\oldstyle{\ixptmit\fam=\@ne}%
     \rm\fi}

%************** 10-point fonts ******************************

\font\xptrm=cmr10
\font\xptmit=cmmi10
\font\xptsy=cmsy10
\font\xptex=cmex10
\font\xptit=cmti10
\font\xptsl=cmsl10
\font\xptbf=cmbx10
\font\xpttt=cmtt10
\font\xptss=cmss10
\font\xptsc=cmcsc10
\font\xptbfs=cmbx10
\font\xptbmit=cmmib10

\skewchar\xptmit='177 \skewchar\xptbmit='177 \skewchar\xptsy='60
\fontdimen16 \xptsy=\the\fontdimen17 \xptsy

\def\xpt{\ifmmode\err@badsizechange\else
     \@mathfontinit
     \textfont0=\xptrm  \scriptfont0=\viiptrm  \scriptscriptfont0=\vptrm
     \textfont1=\xptmit \scriptfont1=\viiptmit \scriptscriptfont1=\vptmit
     \textfont2=\xptsy  \scriptfont2=\viiptsy  \scriptscriptfont2=\vptsy
     \textfont3=\xptex  \scriptfont3=\xptex    \scriptscriptfont3=\xptex
     \textfont\itfam=\xptit
     \scriptfont\itfam=\viiptit
     \scriptscriptfont\itfam=\viiptit
     \textfont\bffam=\xptbf
     \scriptfont\bffam=\viiptbf
     \scriptscriptfont\bffam=\vptbf
     \textfont\bfsfam=\xptbfs
     \scriptfont\bfsfam=\viiptbf
     \scriptscriptfont\bfsfam=\vptbf
     \textfont\bmitfam=\xptbmit
     \scriptfont\bmitfam=\viiptmit
     \scriptscriptfont\bmitfam=\vptmit
     \@fontstyleinit
     \def\rm{\xptrm\fam=\z@}%
     \def\it{\xptit\fam=\itfam}%
     \def\sl{\xptsl}%
     \def\bf{\xptbf\fam=\bffam}%
     \def\tt{\xpttt}%
     \def\ss{\xptss}%
     \def\sc{\xptsc}%
     \def\bfs{\xptbfs\fam=\bfsfam}%
     \def\bmit{\fam=\bmitfam}%
     \def\oldstyle{\xptmit\fam=\@ne}%
     \rm\fi}

%************** Optional 11-point fonts *********************

\def\getxipt{%
     \font\xiptrm=cmr10  scaled\magstephalf
     \font\xiptmit=cmmi10 scaled\magstephalf
     \font\xiptsy=cmsy10 scaled\magstephalf
     \font\xiptex=cmex10 scaled\magstephalf
     \font\xiptit=cmti10 scaled\magstephalf
     \font\xiptsl=cmsl10 scaled\magstephalf
     \font\xiptbf=cmbx10 scaled\magstephalf
     \font\xipttt=cmtt10 scaled\magstephalf
     \font\xiptss=cmss10 scaled\magstephalf
     \skewchar\xiptmit='177 \skewchar\xiptsy='60
     \fontdimen16 \xiptsy=\the\fontdimen17 \xiptsy}

\def\xipt{\ifmmode\err@badsizechange\else
     \@mathfontinit
     \textfont0=\xiptrm  \scriptfont0=\viiiptrm  \scriptscriptfont0=\viptrm
     \textfont1=\xiptmit \scriptfont1=\viiiptmit \scriptscriptfont1=\viptmit
     \textfont2=\xiptsy  \scriptfont2=\viiiptsy  \scriptscriptfont2=\viptsy
     \textfont3=\xiptex  \scriptfont3=\xptex     \scriptscriptfont3=\xptex
     \textfont\itfam=\xiptit
     \scriptfont\itfam=\viiiptit
     \scriptscriptfont\itfam=\viiptit
     \textfont\bffam=\xiptbf
     \scriptfont\bffam=\viiiptbf
     \scriptscriptfont\bffam=\viptbf
     \@fontstyleinit
     \def\rm{\xiptrm\fam=\z@}%
     \def\it{\xiptit\fam=\itfam}%
     \def\sl{\xiptsl}%
     \def\bf{\xiptbf\fam=\bffam}%
     \def\tt{\xipttt}%
     \def\ss{\xiptss}%
     \def\oldstyle{\xiptmit\fam=\@ne}%
     \rm\fi}

%************** 12-point fonts ******************************

\font\xiiptrm=cmr12
\font\xiiptmit=cmmi12
\font\xiiptsy=cmsy10  scaled\magstep1
\font\xiiptex=cmex10  scaled\magstep1
\font\xiiptit=cmti12
\font\xiiptsl=cmsl12
\font\xiiptbf=cmbx12
\font\xiipttt=cmtt12
\font\xiiptss=cmss12
\font\xiiptsc=cmcsc10 scaled\magstep1
\font\xiiptbfs=cmbx10  scaled\magstep1
\font\xiiptbmit=cmmib10 scaled\magstep1

\skewchar\xiiptmit='177 \skewchar\xiiptbmit='177 \skewchar\xiiptsy='60
\fontdimen16 \xiiptsy=\the\fontdimen17 \xiiptsy

\def\xiipt{\ifmmode\err@badsizechange\else
     \@mathfontinit
     \textfont0=\xiiptrm  \scriptfont0=\viiiptrm  \scriptscriptfont0=\viptrm
     \textfont1=\xiiptmit \scriptfont1=\viiiptmit \scriptscriptfont1=\viptmit
     \textfont2=\xiiptsy  \scriptfont2=\viiiptsy  \scriptscriptfont2=\viptsy
     \textfont3=\xiiptex  \scriptfont3=\xptex     \scriptscriptfont3=\xptex
     \textfont\itfam=\xiiptit
     \scriptfont\itfam=\viiiptit
     \scriptscriptfont\itfam=\viiptit
     \textfont\bffam=\xiiptbf
     \scriptfont\bffam=\viiiptbf
     \scriptscriptfont\bffam=\viptbf
     \textfont\bfsfam=\xiiptbfs
     \scriptfont\bfsfam=\viiiptbf
     \scriptscriptfont\bfsfam=\viptbf
     \textfont\bmitfam=\xiiptbmit
     \scriptfont\bmitfam=\viiiptmit
     \scriptscriptfont\bmitfam=\viptmit
     \@fontstyleinit
     \def\rm{\xiiptrm\fam=\z@}%
     \def\it{\xiiptit\fam=\itfam}%
     \def\sl{\xiiptsl}%
     \def\bf{\xiiptbf\fam=\bffam}%
     \def\tt{\xiipttt}%
     \def\ss{\xiiptss}%
     \def\sc{\xiiptsc}%
     \def\bfs{\xiiptbfs\fam=\bfsfam}%
     \def\bmit{\fam=\bmitfam}%
     \def\oldstyle{\xiiptmit\fam=\@ne}%
     \rm\fi}

%************** Optional 13-point fonts *********************

\def\getxiiipt{%
     \font\xiiiptrm=cmr12  scaled\magstephalf
     \font\xiiiptmit=cmmi12 scaled\magstephalf
     \font\xiiiptsy=cmsy9  scaled\magstep2
     \font\xiiiptit=cmti12 scaled\magstephalf
     \font\xiiiptsl=cmsl12 scaled\magstephalf
     \font\xiiiptbf=cmbx12 scaled\magstephalf
     \font\xiiipttt=cmtt12 scaled\magstephalf
     \font\xiiiptss=cmss12 scaled\magstephalf
     \skewchar\xiiiptmit='177 \skewchar\xiiiptsy='60
     \fontdimen16 \xiiiptsy=\the\fontdimen17 \xiiiptsy}

\def\xiiipt{\ifmmode\err@badsizechange\else
     \@mathfontinit
     \textfont0=\xiiiptrm  \scriptfont0=\xptrm  \scriptscriptfont0=\viiptrm
     \textfont1=\xiiiptmit \scriptfont1=\xptmit \scriptscriptfont1=\viiptmit
     \textfont2=\xiiiptsy  \scriptfont2=\xptsy  \scriptscriptfont2=\viiptsy
     \textfont3=\xivptex   \scriptfont3=\xptex  \scriptscriptfont3=\xptex
     \textfont\itfam=\xiiiptit
     \scriptfont\itfam=\xptit
     \scriptscriptfont\itfam=\viiptit
     \textfont\bffam=\xiiiptbf
     \scriptfont\bffam=\xptbf
     \scriptscriptfont\bffam=\viiptbf
     \@fontstyleinit
     \def\rm{\xiiiptrm\fam=\z@}%
     \def\it{\xiiiptit\fam=\itfam}%
     \def\sl{\xiiiptsl}%
     \def\bf{\xiiiptbf\fam=\bffam}%
     \def\tt{\xiiipttt}%
     \def\ss{\xiiiptss}%
     \def\oldstyle{\xiiiptmit\fam=\@ne}%
     \rm\fi}

%************** 14-point fonts ******************************

\font\xivptrm=cmr12   scaled\magstep1
\font\xivptmit=cmmi12  scaled\magstep1
\font\xivptsy=cmsy10  scaled\magstep2
\font\xivptex=cmex10  scaled\magstep2
\font\xivptit=cmti12  scaled\magstep1
\font\xivptsl=cmsl12  scaled\magstep1
\font\xivptbf=cmbx12  scaled\magstep1
\font\xivpttt=cmtt12  scaled\magstep1
\font\xivptss=cmss12  scaled\magstep1
\font\xivptsc=cmcsc10 scaled\magstep2
\font\xivptbfs=cmbx10  scaled\magstep2
\font\xivptbmit=cmmib10 scaled\magstep2

\skewchar\xivptmit='177 \skewchar\xivptbmit='177 \skewchar\xivptsy='60
\fontdimen16 \xivptsy=\the\fontdimen17 \xivptsy

\def\xivpt{\ifmmode\err@badsizechange\else
     \@mathfontinit
     \textfont0=\xivptrm  \scriptfont0=\xptrm  \scriptscriptfont0=\viiptrm
     \textfont1=\xivptmit \scriptfont1=\xptmit \scriptscriptfont1=\viiptmit
     \textfont2=\xivptsy  \scriptfont2=\xptsy  \scriptscriptfont2=\viiptsy
     \textfont3=\xivptex  \scriptfont3=\xptex  \scriptscriptfont3=\xptex
     \textfont\itfam=\xivptit
     \scriptfont\itfam=\xptit
     \scriptscriptfont\itfam=\viiptit
     \textfont\bffam=\xivptbf
     \scriptfont\bffam=\xptbf
     \scriptscriptfont\bffam=\viiptbf
     \textfont\bfsfam=\xivptbfs
     \scriptfont\bfsfam=\xptbfs
     \scriptscriptfont\bfsfam=\viiptbf
     \textfont\bmitfam=\xivptbmit
     \scriptfont\bmitfam=\xptbmit
     \scriptscriptfont\bmitfam=\viiptmit
     \@fontstyleinit
     \def\rm{\xivptrm\fam=\z@}%
     \def\it{\xivptit\fam=\itfam}%
     \def\sl{\xivptsl}%
     \def\bf{\xivptbf\fam=\bffam}%
     \def\tt{\xivpttt}%
     \def\ss{\xivptss}%
     \def\sc{\xivptsc}%
     \def\bfs{\xivptbfs\fam=\bfsfam}%
     \def\bmit{\fam=\bmitfam}%
     \def\oldstyle{\xivptmit\fam=\@ne}%
     \rm\fi}

%************** 17-point fonts ******************************

\font\xviiptrm=cmr17
\font\xviiptmit=cmmi12 scaled\magstep2
\font\xviiptsy=cmsy10 scaled\magstep3
\font\xviiptex=cmex10 scaled\magstep3
\font\xviiptit=cmti12 scaled\magstep2
\font\xviiptbf=cmbx12 scaled\magstep2
\font\xviiptbfs=cmbx10 scaled\magstep3

\skewchar\xviiptmit='177 \skewchar\xviiptsy='60
\fontdimen16 \xviiptsy=\the\fontdimen17 \xviiptsy

\def\xviipt{\ifmmode\err@badsizechange\else
     \@mathfontinit
     \textfont0=\xviiptrm  \scriptfont0=\xiiptrm  \scriptscriptfont0=\viiiptrm
     \textfont1=\xviiptmit \scriptfont1=\xiiptmit \scriptscriptfont1=\viiiptmit
     \textfont2=\xviiptsy  \scriptfont2=\xiiptsy  \scriptscriptfont2=\viiiptsy
     \textfont3=\xviiptex  \scriptfont3=\xiiptex  \scriptscriptfont3=\xptex
     \textfont\itfam=\xviiptit
     \scriptfont\itfam=\xiiptit
     \scriptscriptfont\itfam=\viiiptit
     \textfont\bffam=\xviiptbf
     \scriptfont\bffam=\xiiptbf
     \scriptscriptfont\bffam=\viiiptbf
     \textfont\bfsfam=\xviiptbfs
     \scriptfont\bfsfam=\xiiptbfs
     \scriptscriptfont\bfsfam=\viiiptbf
     \@fontstyleinit
     \def\rm{\xviiptrm\fam=\z@}%
     \def\it{\xviiptit\fam=\itfam}%
     \def\bf{\xviiptbf\fam=\bffam}%
     \def\bfs{\xviiptbfs\fam=\bfsfam}%
     \def\oldstyle{\xviiptmit\fam=\@ne}%
     \rm\fi}

%************** 21-point fonts ******************************

\font\xxiptrm=cmr17  scaled\magstep1
%\font\xxiptmit=cmmi12 scaled\magstep3
%\font\xxiptsy=cmsy10 scaled\magstep4
%\font\xxiptex=cmex10 scaled\magstep4
%\font\xxiptbf=cmbx12 scaled\magstep3

%\skewchar\xxiptmit='177 \skewchar\xxiptsy='60
%\fontdimen16 \xxiptsy=\the\fontdimen17 \xxiptsy

\def\xxipt{\ifmmode\err@badsizechange\else
     \@mathfontinit
%     \textfont0=\xxiptrm  \scriptfont0=\xivptrm  \scriptscriptfont0=\xptrm
%     \textfont1=\xxiptmit \scriptfont1=\xivptmit \scriptscriptfont1=\xptmit
%     \textfont2=\xxiptsy  \scriptfont2=\xivptsy  \scriptscriptfont2=\xptsy
%     \textfont3=\xxiptex  \scriptfont3=\xivptex  \scriptscriptfont3=\xptex
%     \textfont\bffam=\xxiptbf
%     \scriptfont\bffam=\xivptbf
%     \scriptscriptfont\bffam=\xptbf
     \@fontstyleinit
     \def\rm{\xxiptrm\fam=\z@}%
     \rm\fi}

%************** 25-point fonts ******************************

\font\xxvptrm=cmr17  scaled\magstep2
%\font\xxvptmit=cmmi12 scaled\magstep4
%\font\xxvptsy=cmsy10 scaled\magstep5
%\font\xxvptex=cmex10 scaled\magstep5
%\font\xxvptbf=cmbx12 scaled\magstep4

%\skewchar\xxvptmit='177 \skewchar\xxvptsy='60
%\fontdimen16 \xxvptsy=\the\fontdimen17 \xxvptsy

\def\xxvpt{\ifmmode\err@badsizechange\else
     \@mathfontinit
%     \textfont0=\xxvptrm  \scriptfont0=\xviiptrm  \scriptscriptfont0=\xiiptrm
%     \textfont1=\xxvptmit \scriptfont1=\xviiptmit \scriptscriptfont1=\xiiptmit
%     \textfont2=\xxvptsy  \scriptfont2=\xviiptsy  \scriptscriptfont2=\xiiptsy
%     \textfont3=\xxvptex  \scriptfont3=\xviiptex  \scriptscriptfont3=\xiiptex
%     \textfont\bffam=\xxvptbf
%     \scriptfont\bffam=\xviiptbf
%     \scriptscriptfont\bffam=\xiiptbf
     \@fontstyleinit
     \def\rm{\xxvptrm\fam=\z@}%
     \rm\fi}

%************** Other fonts *********************************

%\font\dummy=dummy

%******************************************************************************

\message{Loading jyTeX macros...}

%************************************************************
%*
%*              Simple modifications to plain
%*
%************************************************************
\message{modifications to plain.tex,}

% The "\outer" qualifier is removed from the definitions of \newcount through
% \newif so that they may be used in definitions.  \newif is also changed to
% make \if commands globally defined.

\def\newcount{\alloc@0\count\countdef\insc@unt}
\def\newdimen{\alloc@1\dimen\dimendef\insc@unt}
\def\newskip{\alloc@2\skip\skipdef\insc@unt}
\def\newmuskip{\alloc@3\muskip\muskipdef\@cclvi}
\def\newbox{\alloc@4\box\chardef\insc@unt}
\def\newtoks{\alloc@5\toks\toksdef\@cclvi}
\def\newhelp#1#2{\newtoks#1\global#1\expandafter{\csname#2\endcsname}}
\def\newread{\alloc@6\read\chardef\sixt@@n}
\def\newwrite{\alloc@7\write\chardef\sixt@@n}
\def\newfam{\alloc@8\fam\chardef\sixt@@n}
\def\newinsert#1{\global\advance\insc@unt by\m@ne
     \ch@ck0\insc@unt\count
     \ch@ck1\insc@unt\dimen
     \ch@ck2\insc@unt\skip
     \ch@ck4\insc@unt\box
     \allocationnumber=\insc@unt
     \global\chardef#1=\allocationnumber
     \wlog{\string#1=\string\insert\the\allocationnumber}}
\def\newif#1{\count@\escapechar \escapechar\m@ne
     \expandafter\expandafter\expandafter
          \xdef\@if#1{true}{\let\noexpand#1=\noexpand\iftrue}%
     \expandafter\expandafter\expandafter
          \xdef\@if#1{false}{\let\noexpand#1=\noexpand\iffalse}%
     \global\@if#1{false}\escapechar=\count@}

%************** Some parameter changes **********************

\newlinechar=`\^^J
\overfullrule=0pt

%************** Font-related modifications ******************

% The plain fonts are mapped onto the corresponding jyTeX fonts

% Some control sequences are disabled.

\let\itfam=\undefined

\let\bffam=\undefined

\count18=3

% German sharp s is given a new name (\ss is already taken)

\chardef\sharps="19

% The mathcode assignments of characters in the math italic font are changed to
% allow for switching to boldface.

\mathchardef\alpha="710B
\mathchardef\beta="710C
\mathchardef\gamma="710D
\mathchardef\delta="710E
\mathchardef\epsilon="710F
\mathchardef\zeta="7110
\mathchardef\eta="7111
\mathchardef\theta="7112
\mathchardef\iota="7113
\mathchardef\kappa="7114
\mathchardef\lambda="7115
\mathchardef\mu="7116
\mathchardef\nu="7117
\mathchardef\xi="7118
\mathchardef\pi="7119
\mathchardef\rho="711A
\mathchardef\sigma="711B
\mathchardef\tau="711C
\mathchardef\upsilon="711D
\mathchardef\phi="711E
\mathchardef\chi="711F
\mathchardef\psi="7120
\mathchardef\omega="7121
\mathchardef\varepsilon="7122
\mathchardef\vartheta="7123
\mathchardef\varpi="7124
\mathchardef\varrho="7125
\mathchardef\varsigma="7126
\mathchardef\varphi="7127
\mathchardef\imath="717B
\mathchardef\jmath="717C
\mathchardef\ell="7160
\mathchardef\wp="717D
\mathchardef\partial="7140
\mathchardef\flat="715B
\mathchardef\natural="715C
\mathchardef\sharp="715D

%************** Miscellaneous changes ***********************

% The dimension \p@ (1pt) is replaced with \rp@ (relative pt, defined below),
% whose size is determined by the base type size of the document.

\def\angle{{\vbox{\ialign{$\m@th\scriptstyle##$\crcr
     \not\mathrel{\mkern14mu}\crcr
     \noalign{\nointerlineskip}
     \mkern2.5mu\leaders\hrule height.34\rp@\hfill\mkern2.5mu\crcr}}}}
\def\vdots{\vbox{\baselineskip4\rp@ \lineskiplimit\z@
     \kern6\rp@\hbox{.}\hbox{.}\hbox{.}}}
\def\ddots{\mathinner{\mkern1mu\raise7\rp@\vbox{\kern7\rp@\hbox{.}}\mkern2mu
     \raise4\rp@\hbox{.}\mkern2mu\raise\rp@\hbox{.}\mkern1mu}}
\def\overrightarrow#1{\vbox{\ialign{##\crcr
     \rightarrowfill\crcr
     \noalign{\kern-\rp@\nointerlineskip}
     $\hfil\displaystyle{#1}\hfil$\crcr}}}
\def\overleftarrow#1{\vbox{\ialign{##\crcr
     \leftarrowfill\crcr
     \noalign{\kern-\rp@\nointerlineskip}
     $\hfil\displaystyle{#1}\hfil$\crcr}}}
\def\overbrace#1{\mathop{\vbox{\ialign{##\crcr
     \noalign{\kern3\rp@}
     \downbracefill\crcr
     \noalign{\kern3\rp@\nointerlineskip}
     $\hfil\displaystyle{#1}\hfil$\crcr}}}\limits}
\def\underbrace#1{\mathop{\vtop{\ialign{##\crcr
     $\hfil\displaystyle{#1}\hfil$\crcr
     \noalign{\kern3\rp@\nointerlineskip}
     \upbracefill\crcr
     \noalign{\kern3\rp@}}}}\limits}
\def\big#1{{\hbox{$\left#1\vbox to8.5\rp@ {}\right.\n@space$}}}
\def\Big#1{{\hbox{$\left#1\vbox to11.5\rp@ {}\right.\n@space$}}}
\def\bigg#1{{\hbox{$\left#1\vbox to14.5\rp@ {}\right.\n@space$}}}
\def\Bigg#1{{\hbox{$\left#1\vbox to17.5\rp@ {}\right.\n@space$}}}
\def\@vereq#1#2{\lower.5\rp@\vbox{\baselineskip\z@skip\lineskip-.5\rp@
     \ialign{$\m@th#1\hfil##\hfil$\crcr#2\crcr=\crcr}}}
\def\rlh@#1{\vcenter{\hbox{\ooalign{\raise2\rp@
     \hbox{$#1\rightharpoonup$}\crcr
     $#1\leftharpoondown$}}}}
\def\bordermatrix#1{\begingroup\m@th
     \setbox\z@\vbox{%
          \def\cr{\crcr\noalign{\kern2\rp@\global\let\cr\endline}}%
          \ialign{$##$\hfil\kern2\rp@\kern\p@renwd
               &\thinspace\hfil$##$\hfil&&\quad\hfil$##$\hfil\crcr
               \omit\strut\hfil\crcr
               \noalign{\kern-\baselineskip}%
               #1\crcr\omit\strut\cr}}%
     \setbox\tw@\vbox{\unvcopy\z@\global\setbox\@ne\lastbox}%
     \setbox\tw@\hbox{\unhbox\@ne\unskip\global\setbox\@ne\lastbox}%
     \setbox\tw@\hbox{$\kern\wd\@ne\kern-\p@renwd\left(\kern-\wd\@ne
          \global\setbox\@ne\vbox{\box\@ne\kern2\rp@}%
          \vcenter{\kern-\ht\@ne\unvbox\z@\kern-\baselineskip}%
          \,\right)$}%
     \null\;\vbox{\kern\ht\@ne\box\tw@}\endgroup}
\def\endinsert{\egroup
     \if@mid\dimen@\ht\z@
          \advance\dimen@\dp\z@
          \advance\dimen@12\rp@
          \advance\dimen@\pagetotal
          \ifdim\dimen@>\pagegoal\@midfalse\p@gefalse\fi
     \fi
     \if@mid\bigskip\box\z@
          \bigbreak
     \else\insert\topins{\penalty100 \splittopskip\z@skip
               \splitmaxdepth\maxdimen\floatingpenalty\z@
               \ifp@ge\dimen@\dp\z@
                    \vbox to\vsize{\unvbox\z@\kern-\dimen@}%
               \else\box\z@\nobreak\bigskip
               \fi}%
     \fi
     \endgroup}

% \normalbaselines is removed from \cases and \matrix.

\def\cases#1{\left\{\,\vcenter{\m@th
     \ialign{$##\hfil$&\quad##\hfil\crcr#1\crcr}}\right.}
\def\matrix#1{\null\,\vcenter{\m@th
     \ialign{\hfil$##$\hfil&&\quad\hfil$##$\hfil\crcr
          \mathstrut\crcr
          \noalign{\kern-\baselineskip}
          #1\crcr
          \mathstrut\crcr
          \noalign{\kern-\baselineskip}}}\,}

% \raggedbottom modified slightly

\newif\ifraggedbottom

\def\raggedbottom{\ifraggedbottom\else
     \advance\topskip by\z@ plus60pt \raggedbottomtrue\fi}%
\def\normalbottom{\ifraggedbottom
     \advance\topskip by\z@ plus-60pt \raggedbottomfalse\fi}

%************************************************************
%*
%*              Miscellaneous definitions
%*
%************************************************************
\message{hacks,}

%************** Hack registers ******************************

\toksdef\toks@i=1
\toksdef\toks@ii=2

%************** Basic macros ********************************

\def\TeX{T\kern-.1667em \lower.5ex \hbox{E}\kern-.125em X\null}
\def\jyTeX{{\leavevmode
     \raise.587ex \hbox{\it\j}\kern-.1em \lower.048ex \hbox{\it y}\kern-.12em
     \TeX}}

\let\then=\iftrue
\def\ifnoarg#1\then{\def\hack@{#1}\ifx\hack@\empty}
\def\ifundefined#1\then{%
     \expandafter\ifx\csname\expandafter\blank\string#1\endcsname\relax}
\def\useif#1\then{\csname#1\endcsname}
\def\usename#1{\csname#1\endcsname}
\def\useafter#1#2{\expandafter#1\csname#2\endcsname}

% Modify so that I can have \loop's within \loop's?
\long\def\loop#1\repeat{\def\@iterate{#1\expandafter\@iterate\fi}\@iterate
     \let\@iterate=\relax}
%\long\def\loop#1\repeat{\def\@loopbody{#1}\@iterate}
%\def\@iterate{\@loopbody\let\next=\@iterate\else\let\next=\relax\fi\next}

\let\TeXend=\end
\def\begin#1{\begingroup\def\@@blockname{#1}\usename{begin#1}}
\def\end#1{\usename{end#1}\def\hack@{#1}%
     \ifx\@@blockname\hack@
          \endgroup
     \else\err@badgroup\hack@\@@blockname
     \fi}
\def\@@blockname{}

\def\defaultoption[#1]#2{%
     \def\hack@{\ifx\hack@ii[\toks@={#2}\else\toks@={#2[#1]}\fi\the\toks@}%
     \futurelet\hack@ii\hack@}

\def\markup#1{\let\@@marksf=\empty
     \ifhmode\edef\@@marksf{\spacefactor=\the\spacefactor\relax}\/\fi
     ${}^{\hbox{\subscriptfonts#1}}$\@@marksf}

%************** Time registers ******************************

\newtoks\shortyear
\newtoks\militaryhour
\newtoks\standardhour
\newtoks\minute
\newtoks\amorpm

\def\settime{\count@=\time\divide\count@ by60
     \militaryhour=\expandafter{\number\count@}%
     {\multiply\count@ by-60 \advance\count@ by\time
          \xdef\hack@{\ifnum\count@<10 0\fi\number\count@}}%
     \minute=\expandafter{\hack@}%
     \ifnum\count@<12
          \amorpm={am}
     \else\amorpm={pm}
          \ifnum\count@>12 \advance\count@ by-12 \fi
     \fi
     \standardhour=\expandafter{\number\count@}%
     \def\hack@19##1##2{\shortyear={##1##2}}%
          \expandafter\hack@\the\year}

\def\monthword#1{%
     \ifcase#1
          $\bullet$\err@badcountervalue{monthword}%
          \or January\or February\or March\or April\or May\or June%
          \or July\or August\or September\or October\or November\or December%
     \else$\bullet$\err@badcountervalue{monthword}%
     \fi}

\def\monthabbr#1{%
     \ifcase#1
          $\bullet$\err@badcountervalue{monthabbr}%
          \or Jan\or Feb\or Mar\or Apr\or May\or Jun%
          \or Jul\or Aug\or Sep\or Oct\or Nov\or Dec%
     \else$\bullet$\err@badcountervalue{monthabbr}%
     \fi}

\def\militarytime{\the\militaryhour:\the\minute}
\def\standardtime{\the\standardhour:\the\minute}

%************** Number styles *******************************

\def\@setnumstyle#1#2{\expandafter\global\expandafter\expandafter
     \expandafter\let\expandafter\expandafter
     \csname @\expandafter\blank\string#1style\endcsname
     \csname#2\endcsname}
\def\numstyle#1{\usename{@\expandafter\blank\string#1style}#1}
\def\ifblank#1\then{\useafter\ifx{@\expandafter\blank\string#1}\blank}

\def\blank#1{}

\def\Roman#1{\expandafter\uppercase\expandafter{\romannumeral#1}}
\def\alphabetic#1{%
     \ifcase#1
          $\bullet$\err@badcountervalue{alphabetic}%
          \or a\or b\or c\or d\or e\or f\or g\or h\or i\or j\or k\or l\or m%
          \or n\or o\or p\or q\or r\or s\or t\or u\or v\or w\or x\or y\or z%
     \else$\bullet$\err@badcountervalue{alphabetic}%
     \fi}
\def\Alphabetic#1{\expandafter\uppercase\expandafter{\alphabetic{#1}}}
\def\symbols#1{%
     \ifcase#1
          $\bullet$\err@badcountervalue{symbols}%
          \or*\or\dag\or\ddag\or\S\or$\|$%
          \or**\or\dag\dag\or\ddag\ddag\or\S\S\or$\|\|$%
     \else$\bullet$\err@badcountervalue{symbols}%
     \fi}

%************** String macros *******************************

\catcode`\^^?=13 \def^^?{\relax}

\def\trimleading#1\to#2{\edef#2{#1}%
     \expandafter\@trimleading\expandafter#2#2^^?^^?}
\def\@trimleading#1#2#3^^?{\ifx#2^^?\def#1{}\else\def#1{#2#3}\fi}

\def\trimtrailing#1\to#2{\edef#2{#1}%
     \expandafter\@trimtrailing\expandafter#2#2^^? ^^?\relax}
\def\@trimtrailing#1#2 ^^?#3{\ifx#3\relax\toks@={}%
     \else\def#1{#2}\toks@={\trimtrailing#1\to#1}\fi
     \the\toks@}

\def\trim#1\to#2{\trimleading#1\to#2\trimtrailing#2\to#2}

\catcode`\^^?=15

%************** List macros *********************************

\long\def\additemL#1\to#2{\toks@={\^^\{#1}}\toks@ii=\expandafter{#2}%
     \xdef#2{\the\toks@\the\toks@ii}}

\long\def\additemR#1\to#2{\toks@={\^^\{#1}}\toks@ii=\expandafter{#2}%
     \xdef#2{\the\toks@ii\the\toks@}}

\def\getitemL#1\to#2{\expandafter\@getitemL#1\hack@#1#2}
\def\@getitemL\^^\#1#2\hack@#3#4{\def#4{#1}\def#3{#2}}

%************************************************************
%*
%*             Font-related macros
%*
%************************************************************
\message{font macros,}

%************** Font set-up *********************************

\newdimen\rp@
\newcount\@@sizeindex \@@sizeindex=0
\newcount\@@factori
\newcount\@@factorii
\newcount\@@factoriii
\newcount\@@factoriv

\countdef\maxfam=18
\newfam\itfam
\newfam\bffam
\newfam\bfsfam
\newfam\bmitfam

\def\@mathfontinit{\count@=4
     \loop\textfont\count@=\nullfont
          \scriptfont\count@=\nullfont
          \scriptscriptfont\count@=\nullfont
          \ifnum\count@<\maxfam\advance\count@ by\@ne
     \repeat}

\def\@fontstyleinit{%
     \def\it{\err@fontnotavailable\it}%
     \def\bf{\err@fontnotavailable\bf}%
     \def\bfs{\err@bfstobf}%
     \def\bmit{\err@fontnotavailable\bmit}%
     \def\sc{\err@fontnotavailable\sc}%
     \def\sl{\err@sltoit}%
     \def\ss{\err@fontnotavailable\ss}%
     \def\tt{\err@fontnotavailable\tt}}

\def\@parameterinit#1{\rm\rp@=.1em \@getscaling{#1}%
     \let\^^\=\@doscaling\scalingskipslist
     \setbox\strutbox=\hbox{\vrule
          height.708\baselineskip depth.292\baselineskip width\z@}}

\def\@getfactor#1#2#3#4{\@@factori=#1 \@@factorii=#2
     \@@factoriii=#3 \@@factoriv=#4}

\def\@getscaling#1{\count@=#1 \advance\count@ by-\@@sizeindex\@@sizeindex=#1
     \ifnum\count@<0
          \let\@mulordiv=\divide
          \let\@divormul=\multiply
          \multiply\count@ by\m@ne
     \else\let\@mulordiv=\multiply
          \let\@divormul=\divide
     \fi
     \edef\@@scratcha{\ifcase\count@                {1}{1}{1}{1}\or
          {1}{7}{23}{3}\or     {2}{5}{3}{1}\or      {9}{89}{13}{1}\or
          {6}{25}{6}{1}\or     {8}{71}{14}{1}\or    {6}{25}{36}{5}\or
          {1}{7}{53}{4}\or     {12}{125}{108}{5}\or {3}{14}{53}{5}\or
          {6}{41}{17}{1}\or    {13}{31}{13}{2}\or   {9}{107}{71}{2}\or
          {11}{139}{124}{3}\or {1}{6}{43}{2}\or     {10}{107}{42}{1}\or
          {1}{5}{43}{2}\or     {5}{69}{65}{1}\or    {11}{97}{91}{2}\fi}%
     \expandafter\@getfactor\@@scratcha}

\def\@doscaling#1{\@mulordiv#1by\@@factori\@divormul#1by\@@factorii
     \@mulordiv#1by\@@factoriii\@divormul#1by\@@factoriv}

%************* Size-changing commands ***********************

\newskip\headskip
\newskip\footskip

\def\typesize=#1pt{\count@=#1 \advance\count@ by-10
     \ifcase\count@
          \@setsizex\or\err@badtypesize\or
          \@setsizexii\or\err@badtypesize\or
          \@setsizexiv
     \else\err@badtypesize
     \fi}

\def\@setsizex{\getixpt
     \def\subsubscriptfonts{\vpt}%
          \def\subsubscriptsize{\vpt\@parameterinit{-8}}%
     \def\subscriptfonts{\viipt}\def\subscriptsize{\viipt\@parameterinit{-4}}%
     \def\footnotefonts{\viiipt}\def\footnotesize{\viiipt\@parameterinit{-2}}%
     \def\smallfonts{\ixpt}\def\smallsize{\ixpt\@parameterinit{-1}}%
     \def\normalfonts{\xpt}\def\normalsize{\xpt\@parameterinit{0}}%
     \def\bigfonts{\xiipt}\def\bigsize{\xiipt\@parameterinit{2}}%
     \def\Bigfonts{\xivpt}\def\Bigsize{\xivpt\@parameterinit{4}}%
     \def\biggfonts{\xviipt}\def\biggsize{\xviipt\@parameterinit{6}}%
     \def\Biggfonts{\xxipt}\def\Biggsize{\xxipt\@parameterinit{8}}%
     \def\tinyfonts{\vpt}\def\tinysize{\vpt\@parameterinit{-8}}%
     \def\HUGEFONTS{\xxvpt}\def\HUGESIZE{\xxvpt\@parameterinit{10}}%
     \normalsize\fixedskipslist}

\def\@setsizexii{\getxipt
     \def\subsubscriptfonts{\vipt}%
          \def\subsubscriptsize{\vipt\@parameterinit{-6}}%
     \def\subscriptfonts{\viiipt}%
          \def\subscriptsize{\viiipt\@parameterinit{-2}}%
     \def\footnotefonts{\xpt}\def\footnotesize{\xpt\@parameterinit{0}}%
     \def\smallfonts{\xipt}\def\smallsize{\xipt\@parameterinit{1}}%
     \def\normalfonts{\xiipt}\def\normalsize{\xiipt\@parameterinit{2}}%
     \def\bigfonts{\xivpt}\def\bigsize{\xivpt\@parameterinit{4}}%
     \def\Bigfonts{\xviipt}\def\Bigsize{\xviipt\@parameterinit{6}}%
     \def\biggfonts{\xxipt}\def\biggsize{\xxipt\@parameterinit{8}}%
     \def\Biggfonts{\xxvpt}\def\Biggsize{\xxvpt\@parameterinit{10}}%
     \def\tinyfonts{\vpt}\def\tinysize{\vpt\@parameterinit{-8}}%
     \def\HUGEFONTS{\xxvpt}\def\HUGESIZE{\xxvpt\@parameterinit{10}}%
     \normalsize\fixedskipslist}

\def\@setsizexiv{\getxiiipt
     \def\subsubscriptfonts{\viipt}%
          \def\subsubscriptsize{\viipt\@parameterinit{-4}}%
     \def\subscriptfonts{\xpt}\def\subscriptsize{\xpt\@parameterinit{0}}%
     \def\footnotefonts{\xiipt}\def\footnotesize{\xiipt\@parameterinit{2}}%
     \def\smallfonts{\xiiipt}\def\smallsize{\xiiipt\@parameterinit{3}}%
     \def\normalfonts{\xivpt}\def\normalsize{\xivpt\@parameterinit{4}}%
     \def\bigfonts{\xviipt}\def\bigsize{\xviipt\@parameterinit{6}}%
     \def\Bigfonts{\xxipt}\def\Bigsize{\xxipt\@parameterinit{8}}%
     \def\biggfonts{\xxvpt}\def\biggsize{\xxvpt\@parameterinit{10}}%
     \def\Biggfonts{\err@sizetoolarge\Biggfonts\HUGEFONTS}%
          \def\Biggsize{\err@sizetoolarge\Biggsize\HUGESIZE}%
     \def\tinyfonts{\vpt}\def\tinysize{\vpt\@parameterinit{-8}}%
     \def\HUGEFONTS{\xxvpt}\def\HUGESIZE{\xxvpt\@parameterinit{10}}%
     \normalsize\fixedskipslist}

\def\subsubscriptfonts{\vpt} \def\subsubscriptsize{\vpt\@parameterinit{-8}}
\def\subscriptfonts{\viipt}  \def\subscriptsize{\viipt\@parameterinit{-4}}
\def\footnotefonts{\viiipt}  \def\footnotesize{\viiipt\@parameterinit{-2}}
\def\smallfonts{\err@sizenotavailable\smallfonts}
                             \def\smallsize{\ixpt\@parameterinit{-1}}
\def\normalfonts{\xpt}       \def\normalsize{\xpt\@parameterinit{0}}
\def\bigfonts{\xiipt}        \def\bigsize{\xiipt\@parameterinit{2}}
\def\Bigfonts{\xivpt}        \def\Bigsize{\xivpt\@parameterinit{4}}
\def\biggfonts{\xviipt}      \def\biggsize{\xviipt\@parameterinit{6}}
\def\Biggfonts{\xxipt}       \def\Biggsize{\xxipt\@parameterinit{8}}
\def\tinyfonts{\vpt}         \def\tinysize{\vpt\@parameterinit{-8}}
\def\HUGEFONTS{\xxvpt}       \def\HUGESIZE{\xxvpt\@parameterinit{10}}

%************************************************************
%*
%*             Document layout
%*
%************************************************************
\message{document layout,}

%************** Page format *********************************

\newtoks\everyoutput \everyoutput={}
\newdimen\depthofpage
\newcount\pagenum \pagenum=0

\newdimen\oddtopmargin  \newdimen\eventopmargin
\newdimen\oddleftmargin \newdimen\evenleftmargin
\newtoks\oddhead        \newtoks\evenhead
\newtoks\oddfoot        \newtoks\evenfoot

\def\topmargin{\afterassignment\@seteventop\oddtopmargin}
\def\leftmargin{\afterassignment\@setevenleft\oddleftmargin}
\def\head{\afterassignment\@setevenhead\oddhead}
\def\foot{\afterassignment\@setevenfoot\oddfoot}

\def\@seteventop{\eventopmargin=\oddtopmargin}
\def\@setevenleft{\evenleftmargin=\oddleftmargin}
\def\@setevenhead{\evenhead=\oddhead}
\def\@setevenfoot{\evenfoot=\oddfoot}

\def\pagenumstyle#1{\@setnumstyle\pagenum{#1}}

\newif\ifdraft
\def\draft{\drafttrue\leftmargin=.5in \overfullrule=5pt }

\newif\ifnumup

\def\outputstyle#1{\global\expandafter\let\expandafter
          \@outputstyle\csname#1output\endcsname
     \usename{#1setup}}

\output={\@outputstyle}

\def\normaloutput{\the\everyoutput
     \global\advance\pagenum by\@ne
     \ifodd\pagenum
          \voffset=\oddtopmargin \hoffset=\oddleftmargin
     \else\voffset=\eventopmargin \hoffset=\evenleftmargin
     \fi
     \advance\voffset by-1in  \advance\hoffset by-1in
     \count0=\pagenum
     \expandafter\shipout\pagebox
     \ifnum\outputpenalty>-\@MM\else\dosupereject\fi}

\newdimen\fullhsize
\newbox\leftpage
\newcount\leftpagenum
\newcount\outputpagenum \outputpagenum=0
\let\leftorright=L

\def\twoupoutput{\the\everyoutput
     \global\advance\pagenum by\@ne
     \if L\leftorright
          \global\setbox\leftpage=\leftline{\pagebox}%
          \global\leftpagenum=\pagenum
          \global\let\leftorright=R%
     \else\global\advance\outputpagenum by\@ne
          \ifodd\outputpagenum
               \voffset=\oddtopmargin \hoffset=\oddleftmargin
          \else\voffset=\eventopmargin \hoffset=\evenleftmargin
          \fi
          \advance\voffset by-1in  \advance\hoffset by-1in
          \count0=\leftpagenum \count1=\pagenum
          \shipout\vbox{\hbox to\fullhsize
               {\box\leftpage\hfil\leftline{\pagebox}}}%
          \global\let\leftorright=L%
     \fi
     \ifnum\outputpenalty>-\@MM
     \else\dosupereject
          \if R\leftorright
               \globaldefs=\@ne\head={\hfil}\foot={\hfil}\globaldefs=\z@
               \null\newpage
          \fi
     \fi}

\def\pagebox{\vbox{\makeheadline\pagebody\makefootline}}

\def\makeheadline{%
     \vbox to\z@{\baselinestretch=\@m
          \vskip\topskip\vskip-.708\baselineskip\vskip-\headskip
          \line{\vbox to\ht\strutbox{}%
               \ifodd\pagenum\the\oddhead\else\the\evenhead\fi}%
          \vss}%
     \nointerlineskip}

\def\pagebody{\vbox to\vsize{%
     \boxmaxdepth\maxdepth
     \ifvoid\topins\else\unvbox\topins\fi
     \depthofpage=\dp255
     \unvbox255
     \ifraggedbottom\kern-\depthofpage\vfil\fi
     \ifvoid\footins
     \else\vskip\skip\footins
          \footnoterule
          \unvbox\footins
          \vskip-\footnoteskip
     \fi}}

\def\makefootline{\baselineskip=\footskip
     \line{\ifodd\pagenum\the\oddfoot\else\the\evenfoot\fi}}

%************** Sectioning commands *************************

\newskip\abovechapterskip
\newskip\belowchapterskip
\newskip\abovesectionskip
\newskip\belowsectionskip
\newskip\abovesubsectionskip
\newskip\belowsubsectionskip

\def\chapterstyle#1{\global\expandafter\let\expandafter\@chapterstyle
     \csname#1text\endcsname}
\def\sectionstyle#1{\global\expandafter\let\expandafter\@sectionstyle
     \csname#1text\endcsname}
\def\subsectionstyle#1{\global\expandafter\let\expandafter\@subsectionstyle
     \csname#1text\endcsname}

\def\chapter#1{%
     \ifdim\lastskip=17sp \else\chapterbreak\vskip\abovechapterskip\fi
     \@chapterstyle{\ifblank\chapternumstyle\then
          \else\newchapternum=\next\chapternumformat\ \fi#1}%
     \nobreak\vskip\belowchapterskip\vskip17sp }

\def\section#1{%
     \ifdim\lastskip=17sp \else\sectionbreak\vskip\abovesectionskip\fi
     \@sectionstyle{\ifblank\sectionnumstyle\then
          \else\newsectionnum=\next\sectionnumformat\ \fi#1}%
     \nobreak\vskip\belowsectionskip\vskip17sp }

\def\subsection#1{%
     \ifdim\lastskip=17sp \else\subsectionbreak\vskip\abovesubsectionskip\fi
     \@subsectionstyle{\ifblank\subsectionnumstyle\then
          \else\newsubsectionnum=\next\subsectionnumformat\ \fi#1}%
     \nobreak\vskip\belowsubsectionskip\vskip17sp }

%************** Text formatting commands ********************

\let\TeXunderline=\underline
\let\TeXoverline=\overline
\def\underline#1{\relax\ifmmode\TeXunderline{#1}\else
     $\TeXunderline{\hbox{#1}}$\fi}
\def\overline#1{\relax\ifmmode\TeXoverline{#1}\else
     $\TeXoverline{\hbox{#1}}$\fi}

\def\baselinestretch{\afterassignment\@baselinestretch\count@}
\def\@baselinestretch{\baselineskip=\normalbaselineskip
     \divide\baselineskip by\@m\baselineskip=\count@\baselineskip
     \setbox\strutbox=\hbox{\vrule
          height.708\baselineskip depth.292\baselineskip width\z@}%
     \bigskipamount=\the\baselineskip
          plus.25\baselineskip minus.25\baselineskip
     \medskipamount=.5\baselineskip
          plus.125\baselineskip minus.125\baselineskip
     \smallskipamount=.25\baselineskip
          plus.0625\baselineskip minus.0625\baselineskip}

\def\\{\ifhmode\ifnum\lastpenalty=-\@M\else\hfil\penalty-\@M\fi\fi
     \ignorespaces}
\def\newpage{\vfil\break}

\def\lefttext#1{\par{\@text\leftskip=\z@\rightskip=\centering
     \noindent#1\par}}
\def\righttext#1{\par{\@text\leftskip=\centering\rightskip=\z@
     \noindent#1\par}}
\def\centertext#1{\par{\@text\leftskip=\centering\rightskip=\centering
     \noindent#1\par}}
\def\@text{\parindent=\z@ \parfillskip=\z@ \everypar={}%
     \spaceskip=.3333em \xspaceskip=.5em
     \def\\{\ifhmode\ifnum\lastpenalty=-\@M\else\penalty-\@M\fi\fi
          \ignorespaces}}

\def\beginleft{\par\@text\leftskip=\z@ \rightskip=\centering}
     
\def\beginright{\par\@text\leftskip=\centering\rightskip=\z@ }
     
\def\begincenter{\par\@text\leftskip=\centering\rightskip=\centering}

\def\beginnarrow{\defaultoption[\parindent]\@beginnarrow}
\def\@beginnarrow[#1]{\par\advance\leftskip by#1\advance\rightskip by#1}

\begingroup
\catcode`\[=1 \catcode`\{=11
\gdef\beginignore[\endgroup\bgroup
     \catcode`\e=0 \catcode`\\=12 \catcode`\{=11 \catcode`\f=12 \let\or=\relax
     \let\nd{ignor=\fi \let\}=\egroup
     \iffalse}
\endgroup

\long\def\marginnote#1{\leavevmode
     \edef\@marginsf{\spacefactor=\the\spacefactor\relax}%
     \ifdraft\strut\vadjust{%
          \hbox to\z@{\hskip\hsize\hskip.1in
               \vbox to\z@{\vskip-\dp\strutbox
                    \marginnoteformat
                    \vskip-\ht\strutbox
                    \noindent\strut#1\par
                    \vss}%
               \hss}}%
     \fi
     \@marginsf}

%************** The \bye command ****************************

\newtoks\everybye \everybye={\par\vfil}
\outer\def\bye{\the\everybye
     \footnotecheck
     \prelabelcheck
     \streamcheck
     \supereject
     \TeXend}

%************************************************************
%*
%*             Footnotes
%*
%************************************************************
\message{footnotes,}

\newcount\footnotenum \footnotenum=0
\newskip\footnoteskip
\let\@footnotelist=\empty

\def\footnotenumstyle#1{\@setnumstyle\footnotenum{#1}%
     \useafter\ifx{@footnotenumstyle}\symbols
          \global\let\@footup=\empty
     \else\global\let\@footup=\markup
     \fi}

\def\footnote{\footnotecheck\defaultoption[]\@footnote}
\def\@footnote[#1]{\@footnotemark[#1]\@footnotetext}

\def\footnotemark{\defaultoption[]\@footnotemark}
\def\@footnotemark[#1]{\let\@footsf=\empty
     \ifhmode\edef\@footsf{\spacefactor=\the\spacefactor\relax}\/\fi
     \ifnoarg#1\then
          \global\advance\footnotenum by\@ne
          \@footup{\footnotenumformat}%
          \edef\@@foota{\footnotenum=\the\footnotenum\relax}%
          \expandafter\additemR\expandafter\@footup\expandafter
               {\@@foota\footnotenumformat}\to\@footnotelist
          \global\let\@footnotelist=\@footnotelist
     \else\markup{#1}%
          \additemR\markup{#1}\to\@footnotelist
          \global\let\@footnotelist=\@footnotelist
     \fi
     \@footsf}

\def\footnotetext{%
     \ifx\@footnotelist\empty\err@extrafootnotetext\else\@footnotetext\fi}
\def\@footnotetext{%
     \getitemL\@footnotelist\to\@@foota
     \global\let\@footnotelist=\@footnotelist
     \insert\footins\bgroup
     \footnoteformat
     \splittopskip=\ht\strutbox\splitmaxdepth=\dp\strutbox
     \interlinepenalty=\interfootnotelinepenalty\floatingpenalty=\@MM
     \noindent\llap{\@@foota}\strut
     \bgroup\aftergroup\@footnoteend
     \let\@@scratcha=}
\def\@footnoteend{\strut\par\vskip\footnoteskip\egroup}

\def\footnoterule{\normalfonts
     \kern-.3em \hrule width2in height.04em \kern .26em }

\def\footnotecheck{%
     \ifx\@footnotelist\empty
     \else\err@extrafootnotemark
          \global\let\@footnotelist=\empty
     \fi}

%************************************************************
%*
%*             Labelling macros
%*
%************************************************************
\message{labels,}

\let\@@labeldef=\xdef
\newif\if@labelfile
\newwrite\@labelfile
\let\@prelabellist=\empty

\def\label#1#2{\trim#1\to\@@labarg\edef\@@labtext{#2}%
     \edef\@@labname{lab@\@@labarg}%
     \useafter\ifundefined\@@labname\then\else\@yeslab\fi
     \useafter\@@labeldef\@@labname{#2}%
     \ifstreaming
          \expandafter\toks@\expandafter\expandafter\expandafter
               {\csname\@@labname\endcsname}%
          \immediate\write\streamout{\noexpand\label{\@@labarg}{\the\toks@}}%
     \fi}
\def\@yeslab{%
     \useafter\ifundefined{if\@@labname}\then
          \err@labelredef\@@labarg
     \else\useif{if\@@labname}\then
               \err@labelredef\@@labarg
          \else\global\usename{\@@labname true}%
               \useafter\ifundefined{pre\@@labname}\then
               \else\useafter\ifx{pre\@@labname}\@@labtext
                    \else\err@badlabelmatch\@@labarg
                    \fi
               \fi
               \if@labelfile
               \else\global\@labelfiletrue
                    \immediate\write\sixt@@n{--> Creating file \jobname.lab}%
                    \immediate\openout\@labelfile=\jobname.lab
               \fi
               \immediate\write\@labelfile
                    {\noexpand\prelabel{\@@labarg}{\@@labtext}}%
          \fi
     \fi}

\def\putlab#1{\trim#1\to\@@labarg\edef\@@labname{lab@\@@labarg}%
     \useafter\ifundefined\@@labname\then\@nolab\else\usename\@@labname\fi}
\def\@nolab{%
     \useafter\ifundefined{pre\@@labname}\then
          \undefinedlabelformat
          \err@needlabel\@@labarg
          \useafter\xdef\@@labname{\undefinedlabelformat}%
     \else\usename{pre\@@labname}%
          \useafter\xdef\@@labname{\usename{pre\@@labname}}%
     \fi
     \useafter\newif{if\@@labname}%
     \expandafter\additemR\@@labarg\to\@prelabellist}

\def\prelabel#1{\useafter\gdef{prelab@#1}}

\def\ifundefinedlabel#1\then{%
     \expandafter\ifx\csname lab@#1\endcsname\relax}
\def\useiflab#1\then{\csname iflab@#1\endcsname}

\def\prelabelcheck{{%
     \def\^^\##1{\useiflab{##1}\then\else\err@undefinedlabel{##1}\fi}%
     \@prelabellist}}

%************************************************************
%*
%*             Equation numbering
%*
%************************************************************
\message{equation numbering,}

\newcount\chapternum
\newcount\sectionnum
\newcount\subsectionnum
\newcount\equationnum
\newcount\subequationnum
\newcount\figurenum
\newcount\subfigurenum
\newcount\tablenum
\newcount\subtablenum

\newif\if@subeqncount
\newif\if@subfigcount
\newif\if@subtblcount

\def\newchapternum{\newsectionnum=\z@\@resetnum\chapternum}
\def\newsectionnum{\newsubsectionnum=\z@\@resetnum\sectionnum}
\def\newsubsectionnum{\newequationnum=\z@\newfigurenum=\z@\newtablenum=\z@
     \@resetnum\subsectionnum}
\def\newequationnum{\newsubequationnum=\z@\@resetnum\equationnum}
\def\newsubequationnum{\@resetnum\subequationnum}
\def\newfigurenum{\newsubfigurenum=\z@\@resetnum\figurenum}
\def\newsubfigurenum{\@resetnum\subfigurenum}
\def\newtablenum{\newsubtablenum=\z@\@resetnum\tablenum}
\def\newsubtablenum{\@resetnum\subtablenum}

\def\@resetnum#1{\global\advance#1by1 \edef\next{\the#1\relax}\global#1}

\newchapternum=0

\def\chapternumstyle#1{\@setnumstyle\chapternum{#1}}
\def\sectionnumstyle#1{\@setnumstyle\sectionnum{#1}}
\def\subsectionnumstyle#1{\@setnumstyle\subsectionnum{#1}}
\def\equationnumstyle#1{\@setnumstyle\equationnum{#1}}
\def\subequationnumstyle#1{\@setnumstyle\subequationnum{#1}%
     \ifblank\subequationnumstyle\then\global\@subeqncountfalse\fi
     \ignorespaces}
\def\figurenumstyle#1{\@setnumstyle\figurenum{#1}}
\def\subfigurenumstyle#1{\@setnumstyle\subfigurenum{#1}%
     \ifblank\subfigurenumstyle\then\global\@subfigcountfalse\fi
     \ignorespaces}
\def\tablenumstyle#1{\@setnumstyle\tablenum{#1}}
\def\subtablenumstyle#1{\@setnumstyle\subtablenum{#1}%
     \ifblank\subtablenumstyle\then\global\@subtblcountfalse\fi
     \ignorespaces}

\def\eqnlabel#1{%
     \if@subeqncount
          \newsubequationnum=\next
     \else\newequationnum=\next
          \ifblank\subequationnumstyle\then
          \else\global\@subeqncounttrue
               \newsubequationnum=\@ne
          \fi
     \fi
     \label{#1}{\puteqnformat}(\puteqn{#1})%
     \ifdraft\rlap{\hskip.1in{\tt#1}}\fi}

\let\puteqn=\putlab

\def\equation#1#2{\useafter\gdef{eqn@#1}{#2\eqno\eqnlabel{#1}}}
\def\Equation#1{\useafter\gdef{eqn@#1}}

\def\putequation#1{\useafter\ifundefined{eqn@#1}\then
     \err@undefinedeqn{#1}\else\usename{eqn@#1}\fi}

\def\eqnseriesstyle#1{\gdef\@eqnseriesstyle{#1}}
\def\begineqnseries{\subequationnumstyle{\@eqnseriesstyle}%
     \defaultoption[]\@begineqnseries}
\def\@begineqnseries[#1]{\edef\@@eqnname{#1}}
\def\endeqnseries{\subequationnumstyle{blank}%
     \expandafter\ifnoarg\@@eqnname\then
     \else\label\@@eqnname{\puteqnformat}%
     \fi
     \aftergroup\ignorespaces}

\def\figlabel#1{%
     \if@subfigcount
          \newsubfigurenum=\next
     \else\newfigurenum=\next
          \ifblank\subfigurenumstyle\then
          \else\global\@subfigcounttrue
               \newsubfigurenum=\@ne
          \fi
     \fi
     \label{#1}{\putfigformat}\putfig{#1}%
     {\def\marginnoteformat{\tt}\marginnote{#1}}}

\let\putfig=\putlab

\def\figseriesstyle#1{\gdef\@figseriesstyle{#1}}
\def\beginfigseries{\subfigurenumstyle{\@figseriesstyle}%
     \defaultoption[]\@beginfigseries}
\def\@beginfigseries[#1]{\edef\@@figname{#1}}
\def\endfigseries{\subfigurenumstyle{blank}%
     \expandafter\ifnoarg\@@figname\then
     \else\label\@@figname{\putfigformat}%
     \fi
     \aftergroup\ignorespaces}

\def\tbllabel#1{%
     \if@subtblcount
          \newsubtablenum=\next
     \else\newtablenum=\next
          \ifblank\subtablenumstyle\then
          \else\global\@subtblcounttrue
               \newsubtablenum=\@ne
          \fi
     \fi
     \label{#1}{\puttblformat}\puttbl{#1}%
     {\def\marginnoteformat{\tt}\marginnote{#1}}}

\let\puttbl=\putlab

\def\tblseriesstyle#1{\gdef\@tblseriesstyle{#1}}
\def\begintblseries{\subtablenumstyle{\@tblseriesstyle}%
     \defaultoption[]\@begintblseries}
\def\@begintblseries[#1]{\edef\@@tblname{#1}}
\def\endtblseries{\subtablenumstyle{blank}%
     \expandafter\ifnoarg\@@tblname\then
     \else\label\@@tblname{\puttblformat}%
     \fi
     \aftergroup\ignorespaces}

%************************************************************
%*
%*             Reference numbering
%*
%************************************************************
\message{reference numbering,}

\newcount\referencenum \referencenum=0
\newcount\@@prerefcount \@@prerefcount=0
\newcount\@@thisref
\newcount\@@lastref
\newcount\@@loopref
\newcount\@@refseq
\newdimen\refnumindent
\let\@undefreflist=\empty

\def\referencenumstyle#1{\@setnumstyle\referencenum{#1}}

\def\referencestyle#1{\usename{@ref#1}}

\def\@refsequential{%
     \gdef\@refpredef##1{\global\advance\referencenum by\@ne
          \let\^^\=0\label{##1}{\^^\{\the\referencenum}}%
          \useafter\gdef{ref@\the\referencenum}{{##1}{\undefinedlabelformat}}}%
     \gdef\@reference##1##2{%
          \ifundefinedlabel##1\then
          \else\def\^^\####1{\global\@@thisref=####1\relax}\putlab{##1}%
               \useafter\gdef{ref@\the\@@thisref}{{##1}{##2}}%
          \fi}%
     \gdef\endputreferences{%
          \loop\ifnum\@@loopref<\referencenum
                    \advance\@@loopref by\@ne
                    \expandafter\expandafter\expandafter\@printreference
                         \csname ref@\the\@@loopref\endcsname
          \repeat
          \par}}

\def\@refpreordered{%
     \gdef\@refpredef##1{\global\advance\referencenum by\@ne
          \additemR##1\to\@undefreflist}%
     \gdef\@reference##1##2{%
          \ifundefinedlabel##1\then
          \else\global\advance\@@loopref by\@ne
               {\let\^^\=0\label{##1}{\^^\{\the\@@loopref}}}%
               \@printreference{##1}{##2}%
          \fi}
     \gdef\endputreferences{%
          \def\^^\####1{\useiflab{####1}\then
               \else\reference{####1}{\undefinedlabelformat}\fi}%
          \@undefreflist
          \par}}

\def\beginprereferences{\par
     \def\reference##1##2{\global\advance\referencenum by1\@ne
          \let\^^\=0\label{##1}{\^^\{\the\referencenum}}%
          \useafter\gdef{ref@\the\referencenum}{{##1}{##2}}}}
\def\endprereferences{\global\@@prerefcount=\the\referencenum\par}

\def\beginputreferences{\par
     \refnumindent=\z@\@@loopref=\z@
     \loop\ifnum\@@loopref<\referencenum
               \advance\@@loopref by\@ne
               \setbox\z@=\hbox{\referencenum=\@@loopref
                    \referencenumformat\enskip}%
               \ifdim\wd\z@>\refnumindent\refnumindent=\wd\z@\fi
     \repeat
     \putreferenceformat
     \@@loopref=\z@
     \loop\ifnum\@@loopref<\@@prerefcount
               \advance\@@loopref by\@ne
               \expandafter\expandafter\expandafter\@printreference
                    \csname ref@\the\@@loopref\endcsname
     \repeat
     \let\reference=\@reference}

\def\@printreference#1#2{\ifx#2\undefinedlabelformat\err@undefinedref{#1}\fi
     \noindent\ifdraft\rlap{\hskip\hsize\hskip.1in \tt#1}\fi
     \llap{\referencenum=\@@loopref\referencenumformat\enskip}#2\par}

\def\reference#1#2{{\par\refnumindent=\z@\putreferenceformat\noindent#2\par}}

\def\putref#1{\trim#1\to\@@refarg
     \expandafter\ifnoarg\@@refarg\then
          \toks@={\relax}%
     \else\@@lastref=-\@m\def\@@refsep{}\def\@more{\@nextref}%
          \toks@={\@nextref#1,,}%
     \fi\the\toks@}
\def\@nextref#1,{\trim#1\to\@@refarg
     \expandafter\ifnoarg\@@refarg\then
          \let\@more=\relax
     \else\ifundefinedlabel\@@refarg\then
               \expandafter\@refpredef\expandafter{\@@refarg}%
          \fi
          \def\^^\##1{\global\@@thisref=##1\relax}%
          \global\@@thisref=\m@ne
          \setbox\z@=\hbox{\putlab\@@refarg}%
     \fi
     \advance\@@lastref by\@ne
     \ifnum\@@lastref=\@@thisref\advance\@@refseq by\@ne\else\@@refseq=\@ne\fi
     \ifnum\@@lastref<\z@
     \else\ifnum\@@refseq<\thr@@
               \@@refsep\def\@@refsep{,}%
               \ifnum\@@lastref>\z@
                    \advance\@@lastref by\m@ne
                    {\referencenum=\@@lastref\putrefformat}%
               \else\undefinedlabelformat
               \fi
          \else\def\@@refsep{--}%
          \fi
     \fi
     \@@lastref=\@@thisref
     \@more}

%************************************************************
%*
%*             Job streaming
%*
%************************************************************
\message{streaming,}

\newif\ifstreaming

\def\streamto{\defaultoption[\jobname]\@streamto}
\def\@streamto[#1]{\global\streamingtrue
     \immediate\write\sixt@@n{--> Streaming to #1.str}%
     \newwrite\streamout\immediate\openout\streamout=#1.str }

\def\streamfrom{\defaultoption[\jobname]\@streamfrom}
\def\@streamfrom[#1]{\newread\streamin\openin\streamin=#1.str
     \ifeof\streamin
          \expandafter\err@nostream\expandafter{#1.str}%
     \else\immediate\write\sixt@@n{--> Streaming from #1.str}%
          \let\@@labeldef=\gdef
          \ifstreaming
               \edef\@elc{\endlinechar=\the\endlinechar}%
               \endlinechar=\m@ne
               \loop\read\streamin to\@@scratcha
                    \ifeof\streamin
                         \streamingfalse
                    \else\toks@=\expandafter{\@@scratcha}%
                         \immediate\write\streamout{\the\toks@}%
                    \fi
                    \ifstreaming
               \repeat
               \@elc
               \input #1.str
               \streamingtrue
          \else\input #1.str
          \fi
          \let\@@labeldef=\xdef
     \fi}

\def\streamcheck{\ifstreaming
     \immediate\write\streamout{\pagenum=\the\pagenum}%
     \immediate\write\streamout{\footnotenum=\the\footnotenum}%
     \immediate\write\streamout{\referencenum=\the\referencenum}%
     \immediate\write\streamout{\chapternum=\the\chapternum}%
     \immediate\write\streamout{\sectionnum=\the\sectionnum}%
     \immediate\write\streamout{\subsectionnum=\the\subsectionnum}%
     \immediate\write\streamout{\equationnum=\the\equationnum}%
     \immediate\write\streamout{\subequationnum=\the\subequationnum}%
     \immediate\write\streamout{\figurenum=\the\figurenum}%
     \immediate\write\streamout{\subfigurenum=\the\subfigurenum}%
     \immediate\write\streamout{\tablenum=\the\tablenum}%
     \immediate\write\streamout{\subtablenum=\the\subtablenum}%
     \immediate\closeout\streamout
     \fi}

%************************************************************
%*
%*             Error messages
%*
%************************************************************

\def\err@badtypesize{%
     \errhelp={The limited availability of certain fonts requires^^J%
          that the base type size be 10pt, 12pt, or 14pt.^^J}%
     \errmessage{--> Illegal base type size}}

\def\err@badsizechange{\immediate\write\sixt@@n
     {--> Size change not allowed in math mode, ignored}}

\def\err@sizetoolarge#1{\immediate\write\sixt@@n
     {--> \noexpand#1 too big, substituting HUGE}}

\def\err@sizenotavailable#1{\immediate\write\sixt@@n
     {--> Size not available, \noexpand#1 ignored}}

\def\err@fontnotavailable#1{\immediate\write\sixt@@n
     {--> Font not available, \noexpand#1 ignored}}

\def\err@sltoit{\immediate\write\sixt@@n
     {--> Style \noexpand\sl not available, substituting \noexpand\it}%
     \it}

\def\err@bfstobf{\immediate\write\sixt@@n
     {--> Style \noexpand\bfs not available, substituting \noexpand\bf}%
     \bf}

\def\err@badgroup#1#2{%
     \errhelp={The block you have just tried to close was not the one^^J%
          most recently opened.^^J}%
     \errmessage{--> \noexpand\end{#1} doesn't match \noexpand\begin{#2}}}

\def\err@badcountervalue#1{\immediate\write\sixt@@n
     {--> Counter (#1) out of bounds}}

\def\err@extrafootnotemark{\immediate\write\sixt@@n
     {--> \noexpand\footnotemark command
          has no corresponding \noexpand\footnotetext}}

\def\err@extrafootnotetext{%
     \errhelp{You have given a \noexpand\footnotetext command without first
          specifying^^Ja \noexpand\footnotemark.^^J}%
     \errmessage{--> \noexpand\footnotetext command has no corresponding
          \noexpand\footnotemark}}

\def\err@labelredef#1{\immediate\write\sixt@@n
     {--> Label "#1" redefined}}

\def\err@badlabelmatch#1{\immediate\write\sixt@@n
     {--> Definition of label "#1" doesn't match value in \jobname.lab}}

\def\err@needlabel#1{\immediate\write\sixt@@n
     {--> Label "#1" cited before its definition}}

\def\err@undefinedlabel#1{\immediate\write\sixt@@n
     {--> Label "#1" cited but never defined}}

\def\err@undefinedeqn#1{\immediate\write\sixt@@n
     {--> Equation "#1" not defined}}

\def\err@undefinedref#1{\immediate\write\sixt@@n
     {--> Reference "#1" not defined}}

\def\err@nostream#1{%
     \errhelp={You have tried to input a stream file that doesn't exist.^^J}%
     \errmessage{--> Stream file #1 not found}}

%************************************************************
%*
%*             Initialization
%*
%************************************************************
\message{jyTeX initialization}

\everyjob{\immediate\write16{--> jyTeX version \fmtversion}%
     \edef\@@jobname{\jobname}%
%     \openin0=\inputpath jysupp
%     \ifeof0
%     \else\closein0
%          \immediate\write16{--> Additional macros loaded from jysupp.tex}%
%          \jyinput jysupp
%     \fi
%     \openin0=\inputpath jylocal
%     \ifeof0
%     \else\closein0
%          \immediate\write16{--> Additional macros loaded from jylocal.tex}%
%          \jyinput jylocal
%     \fi
     \edef\jobname{\@@jobname}%
     \settime
     \openin0=\jobname.lab
     \ifeof0
     \else\closein0
          \immediate\write16{--> Getting labels from file \jobname.lab}%
          \input\jobname.lab
     \fi}

%************** Spacing *************************************

\def\fixedskipslist{%
     \^^\{\topskip}%
     \^^\{\splittopskip}%
     \^^\{\maxdepth}%
     \^^\{\skip\topins}%
     \^^\{\skip\footins}%
     \^^\{\headskip}%
     \^^\{\footskip}}

\def\scalingskipslist{%
     \^^\{\p@renwd}%
     \^^\{\delimitershortfall}%
     \^^\{\nulldelimiterspace}%
     \^^\{\scriptspace}%
     \^^\{\jot}%
     \^^\{\normalbaselineskip}%
     \^^\{\normallineskip}%
     \^^\{\normallineskiplimit}%
     \^^\{\baselineskip}%
     \^^\{\lineskip}%
     \^^\{\lineskiplimit}%
     \^^\{\bigskipamount}%
     \^^\{\medskipamount}%
     \^^\{\smallskipamount}%
     \^^\{\parskip}%
     \^^\{\parindent}%
     \^^\{\abovedisplayskip}%
     \^^\{\belowdisplayskip}%
     \^^\{\abovedisplayshortskip}%
     \^^\{\belowdisplayshortskip}%
     \^^\{\abovechapterskip}%
     \^^\{\belowchapterskip}%
     \^^\{\abovesectionskip}%
     \^^\{\belowsectionskip}%
     \^^\{\abovesubsectionskip}%
     \^^\{\belowsubsectionskip}}

%************** Document layout *****************************

\def\twoupsetup{%                                % setup for twoup style
     \topmargin=.75in
     \leftmargin=.5in
     \vsize=6.9in
     \hsize=4.75in
     \fullhsize=10in
     \let\draft=\relax}

\outputstyle{normal}                             % page style

\def\marginnoteformat{\subscriptsize             % paragraphing of margin notes
     \hsize=1in \baselinestretch=1000 \everypar={}%
     \tolerance=5000 \hbadness=5000 \parskip=0pt \parindent=0pt
     \leftskip=0pt \rightskip=0pt \raggedright}

%\head={\ifdraft\normalfonts\it\hfil DRAFT\hfil   % format of headline
%     \llap{\number\day\ \monthword\month\militarytime}\else\hfil\fi}

\head={\ifnumup\normalfonts\hfill\numstyle\pagenum  % format of headline
        \else\hfil\fi}

\foot={\ifnumup\hfil\else
\hfil\normalfonts\numstyle\pagenum\hfil\fi}  % format of footline

\normalbaselineskip=12pt                         % usual \baselineskip
\normallineskip=0pt                              % usual \lineskip
\normallineskiplimit=0pt                         % usual \lineskiplimit
\normalbaselines                                 % set \baselineskip

\topskip=.85\baselineskip
\splittopskip=\topskip
\headskip=2\baselineskip
\footskip=\headskip

\pagenumstyle{arabic}                            % counter style

\parskip=0pt                                     % no skip between paragraphs
\parindent=20pt                                  % usual \parindent

\baselinestretch=1000                            % set \big-, \med-, \smallskip

%************** Sectioning **********************************

\chapterstyle{left}                              % position of heading
\chapternumstyle{blank}                          % counter style
\def\chapterbreak{\newpage}                      % break before heading
\abovechapterskip=0pt                            % space before heading
\belowchapterskip=1.5\baselineskip               % space after heading
     plus.38\baselineskip minus.38\baselineskip
\def\chapternumformat{\numstyle\chapternum.}     % format of heading counter

\sectionstyle{left}                              % position of heading
\sectionnumstyle{blank}                          % counter style
\def\sectionbreak{\vskip0pt plus4\baselineskip\penalty-100
     \vskip0pt plus-4\baselineskip}              % break before heading
\abovesectionskip=1.5\baselineskip               % space before heading
     plus.38\baselineskip minus.38\baselineskip
\belowsectionskip=\the\baselineskip              % space after heading
     plus.25\baselineskip minus.25\baselineskip
\def\sectionnumformat{%                          % format of heading counter
     \ifblank\chapternumstyle\then\else\numstyle\chapternum.\fi
     \numstyle\sectionnum.}

\subsectionstyle{left}                           % position of heading
\subsectionnumstyle{blank}                       % counter style
\def\subsectionbreak{\vskip0pt plus4\baselineskip\penalty-100
     \vskip0pt plus-4\baselineskip}              % break before heading
\abovesubsectionskip=\the\baselineskip           % space before heading
     plus.25\baselineskip minus.25\baselineskip
\belowsubsectionskip=.75\baselineskip            % space after heading
     plus.19\baselineskip minus.19\baselineskip
\def\subsectionnumformat{%                       % format of heading counter
     \ifblank\chapternumstyle\then\else\numstyle\chapternum.\fi
     \ifblank\sectionnumstyle\then\else\numstyle\sectionnum.\fi
     \numstyle\subsectionnum.}

%************** Footnotes ***********************************

\footnotenumstyle{symbols}                       % counter style
\footnoteskip=0pt                                % jyTeX spacing parameter
\def\footnotenumformat{\numstyle\footnotenum}    % \footnotemark format
\def\footnoteformat{\footnotesize                % paragraphing of text
     \everypar={}\parskip=0pt \parfillskip=0pt plus1fil
     \leftskip=1em \rightskip=0pt
     \spaceskip=0pt \xspaceskip=0pt
     \def\\{\ifhmode\ifnum\lastpenalty=-10000
          \else\hfil\penalty-10000 \fi\fi\ignorespaces}}

%************** Labels **************************************

\def\undefinedlabelformat{$\bullet$}             % mark for undefined label

%************** Equation numbering **************************

\equationnumstyle{arabic}                        % counter style
\subequationnumstyle{blank}                      % counter style
\figurenumstyle{arabic}                          % counter style
\subfigurenumstyle{blank}                        % counter style
\tablenumstyle{arabic}                           % counter style
\subtablenumstyle{blank}                         % counter style

\eqnseriesstyle{alphabetic}                      % sub-counter style for series
\figseriesstyle{alphabetic}                      % sub-counter style for series
\tblseriesstyle{alphabetic}                      % sub-counter style for series

\def\puteqnformat{\hbox{%                        % equation number format
     \ifblank\chapternumstyle\then\else\numstyle\chapternum.\fi
     \ifblank\sectionnumstyle\then\else\numstyle\sectionnum.\fi
     \ifblank\subsectionnumstyle\then\else\numstyle\subsectionnum.\fi
     \numstyle\equationnum
     \numstyle\subequationnum}}
\def\putfigformat{\hbox{%                        % figure number format
     \ifblank\chapternumstyle\then\else\numstyle\chapternum.\fi
     \ifblank\sectionnumstyle\then\else\numstyle\sectionnum.\fi
     \ifblank\subsectionnumstyle\then\else\numstyle\subsectionnum.\fi
     \numstyle\figurenum
     \numstyle\subfigurenum}}
\def\puttblformat{\hbox{%                        % table number format
     \ifblank\chapternumstyle\then\else\numstyle\chapternum.\fi
     \ifblank\sectionnumstyle\then\else\numstyle\sectionnum.\fi
     \ifblank\subsectionnumstyle\then\else\numstyle\subsectionnum.\fi
     \numstyle\tablenum
     \numstyle\subtablenum}}

%************** Reference numbering *************************

\referencestyle{sequential}                      % referencing method
\referencenumstyle{arabic}                       % counter style
\def\putrefformat{\numstyle\referencenum}        % format of reference citation
\def\referencenumformat{\numstyle\referencenum.} % format of number in list
\def\putreferenceformat{%                        % paragraphing of list
     \everypar={\hangindent=1em \hangafter=1 }%
     \def\\{\hfil\break\null\hskip-1em \ignorespaces}%
     \leftskip=\refnumindent\parindent=0pt \interlinepenalty=1000 }

%************** Font initialization *************************

\normalsize

%******************************************************************************

\def\fmtversion{2.6M (June 1992)}
%%%%%%%%%%%%%%%%%%%%%%%%%%%%%%%%%%%%%%%%%%%%%%%%%%%%%%%%%%%%%%%%%%%%%
%
%%%%%%%%%%%%%%%%%%%%%%%%%%%%%%%%%%%%%%%%%%%%%%%%%%%%%%%%%%%%%%%%%%%%%
%
% Finally some useful macros for journals, ...
%
\def\JSP#1#2#3{{\sl J. Stat. Phys.} {\bf #1} (#2) #3}
\def\PRL#1#2#3{{\sl Phys. Rev. Lett.} {\bf#1} (#2) #3}

\def\NPB#1#2#3{{\sl Nucl. Phys.} {\bf B#1} (#2) #3}

\def\PRB#1#2#3{{\sl Phys. Rev.} {\bf B#1} (#2) #3}

\def\PLA#1#2#3{{\sl Phys. Lett.} {\bf #1A} (#2) #3}
\def\JMP#1#2#3{{\sl J. Math. Phys.} {\bf #1} (#2) #3}

\def\PTP#1#2#3{{\sl Prog. Theor. Phys.} {\bf #1} (#2) #3}

\def\RMP#1#2#3{{\sl Rev. Mod. Phys.} {\bf #1} (#2) #3}

\def\IJMPB#1#2#3{{\sl Int. J. Mod. Phys.} {\bf B#1} (#2) #3}

\def\AdP#1#2#3{{\sl Advances in Phys.} {\bf #1} (#2) #3}

\def\JPA#1#2#3{{\sl J. Physics} {\bf A#1} (#2) #3}
\def\JPC#1#2#3{{\sl J. Physics} {\bf C#1} (#2) #3}
\def\JPCM#1#2#3{{\sl J. Physics Cond. Mat.} {\bf #1} (#2) #3}
\def\JPSJ#1#2#3{{\sl J. Phys. Soc. Japan} {\bf #1} (#2) #3}

\def\JETP#1#2#3{{\sl Sov. Phys. JETP} {\bf #1} (#2) #3}
\def\JETPL#1#2#3{{\sl Sov. Phys. JETP Lett.} {\bf #1} (#2) #3}

\def\ZPB#1#2#3{{\sl Z.Phys.} {\bf B#1} (#2) #3}
\def\sci#1#2#3{{\sl Science} {\bf #1} (#2) #3}

\def\upref#1/{\markup{[\putref{#1}]}}
\def\Idoubled#1{{\rm I\kern-.22em #1}}
\def\Odoubled#1{{\setbox0=\hbox{\rm#1}%
     \dimen@=\ht0 \dimen@ii=.04em \advance\dimen@ by-\dimen@ii
     \rlap{\kern.26em \vrule height\dimen@ depth-\dimen@ii width.075em}\box0}}

\catcode`\@=12

\typesize=12pt
\footnotenumstyle{arabic}
\baselineskip=16pt
\null\vskip 3cm
\begin{center}
{\bigfonts\sc Exactly Solvable Models \\}
\smallskip
{\bigfonts\sc of Strongly Correlated Electrons}
\vskip 2cm
{\smallfonts\rm Dedicated to Prof. C.N. Yang on the occasion of his
seventieth birthday}

\vskip 2cm
Editors :
\vskip 1cm
{\ss Vladimir E. Korepin}
\smallskip
and
\smallskip
{\ss Fabian H.L. E\char'31ler}
\vskip .5cm
{\it Institute for Theoretical Physics\vskip 4pt
State University of New York at Stony Brook\vskip 4pt
Stony Brook, NY~~11794-3840}
\vskip .3cm
\end{center}
\vfil\break
\vfill\newpage

\section{\sc Introduction}
In recent years there has been a strong increase of interest in
strongly correlated electronic systems in the Condensed Matter
community. The reason for this development is that certain phenomena
like high-$T_c$ superconductivity cannot seem to be explained in the
framework of weak interactions, that is mean field theory and
perturbation around it. There are mainly two different approaches in
dealing with strong correlations : numerical studies and exact
analytical solutions. Analytical solutions provide a complete and
unambiguous picture of the dynamics of the models under consideration.
The analytical approach is mainly
limited to one spatial dimension, but there are indications that
two-dimensional systems may share certain features with their
one-dimensional analogs (see {\sl e.g.} P.W. Anderson in
[\putref{pw}]; for a modified Hubbard model it was recently
shown that the ground state structure is identical in one and two
dimensions\upref eks5/).
This volume is devoted to exact solutions of models of strongly
correlated electrons in one spatial dimension by means of the Bethe
Ansatz.\par
The first such exactly solved model is the non-relativistic continuum
model of electrons with local interaction, solved by C.N. Yang
in 1967\upref ya1/ (see also [\putref{Gau2}]). In the solution of the
model the nested Bethe Ansatz, which is the basis for all exact
solutions of electronic models in one dimension, was discovered.\\
The most important model of strongly correlated electrons, and the
central topic of this volume, is the Hubbard model. The first chapter
is devoted to its exact solution in one space and one time
dimension, which started with E.H. Lieb's and F.Y. Wu's work of 1968.
The chapter starts with reprints dealing with the construction of
eigenstates of the Hubbard hamiltonian and the determination of the
ground state and excitation spectrum at zero temperature. The next few
topics are the study of magnetic properties in an external magnetic
field, the thermodynamics of the model, and the asymptotical behaviour
of correlation functions. The chapter closes with reprints on
transport properties of the model.\par
The second chapter is devoted to the $t$-$J$ model at the
supersymmetric point $J=\pm 2t$.
The $t$-$J$ model first appeared as an effective model describing the strong
coupling limit of the Hubbard model (for $J\ll t$). Later the model
was reinvented to describe copper-oxide planes in high-$T_c$
superconductors. Nowadays the $t$-$J$ model is the most thoroughly
studied model of strongly correlated electrons after the Hubbard
model. The organization of chapter II is similar to that of chapter
I.\\
The third chapter is about other solvable models of strongly
correlated electrons. The first half of the chapter is devoted to
various electronic models with local interaction - the nonrelativistic
continuum model of electrons with local interaction, the Luttinger
model, and models which were constructed recently in relation with
high-$T_c$ superconductivity.
The second part of the chapter deals with models with long-range
interactions, which have recently attracted much attention in relation
with fractional statistics.\par
The reprinted papers are complemented by a list of some 230 references
at the end of the volume. In this list reprinted papers are marked by
a $\bullet$. When referring to reprinted papers in the comments we
first give the chapter number and then the number of the reprint
within the chapter, {\sl e.g.} {\ss [repr.III.A.2]} would refer to the
paper by M. Takahashi reprinted in part A of chapter III under the
number 2.\par

The central topic of this volume are lattice models of strongly
correlated electrons. We did not include reprints on models describing
interactions of electrons with impurities (see the excellent reviews
[\putref{wiegm,low}]) or other excitations in this volume. We also did
not include papers dealing with purely mathematical aspects of
integrable models. There are many excellent books dealing with these
issues\upref baxter, Gau, vladb, is, fadtak, jimbo, izs, liebmat, mw,
shas/ and we refer the interested reader to them. We also did not
include any reviews on the Hubbard model.\par
Due to financial problems caused by republication fees charged by
certain journals we were not able to include a significant number of
papers which we had originally planned to get reprinted.
We are very sorry about this unfortunate development but are confident
that the volume in its current form will still be helpful for
researchers and students.\par
While working on this volume we have benefitted greatly from
discussions with L.D. Faddeev, F.D.M. Haldane, A.I. Larkin, E.H. Lieb,
B.S. Shastry, E.K. Sklyanin, B. Sutherland, F. Woynarovich, and C.N.
Yang. We would like to take this  opportunity to sincerely thank them
for their help and support.
\vskip .5cm
Stony Brook, July 1993
\vskip 1cm
Vladimir E. Korepin\qquad\qquad Fabian H.L. E\char'31ler
\vfill\newpage

\section{I. \sc Reprints on the One-Dimensional Hubbard Model}
\item{1.} E.H. Lieb and F.Y. Wu :\ {\sl Absence of Mott Transition in
an Exact Solution of the Short-Range One-Band Model in One
Dimension},\ \PRL{20}{1968}{1445}.
\smallskip
\item{2.} A.A. Ovchinnikov :\ \hskip -2pt{\sl Excitation Spectrum of the
One-Dimensional Hubbard Model},\ \JETP{30}{1970}{1160}.
\smallskip
\item{3.} C.F. Coll :\ {\sl Excitation spectrum in the one-dimensional
Hubbard model},\ \PRB{9}{1974}{2150}.
\smallskip
\item{4.} F. Woynarovich :\ {\sl Excitations with complex wavenumbers
in a Hubbard chain : I. States with one pair of complex wavenumbers},\
\JPC{15}{1982}{85}.
\smallskip
\item{5.} F. Woynarovich :\ {\sl Excitations with complex wavenumbers
in a Hubbard chain : II. States with several pairs of complex
wavenumbers},\ \JPC{15}{1982}{97}.
\smallskip
\item{6.} F. Woynarovich :\ {\sl Spin Excitations in a Hubbard
chain},\ \JPC{16}{1983}{5293}.
\smallskip
\item{7.} F. Woynarovich :\ {\sl Low-energy excited states in a
Hubbard chain with on-site attraction},\ \JPC{16}{1983}{6593}.
\smallskip
\item{8.} A. Kl\"umper, A. Schadschneider and J. Zittartz :\ {\sl A new
method for the excitations of the one-dimensional Hubbard model},\
\ZPB{78}{1990}{99}.
\smallskip
\item{9.} M. Takahashi :\ {\sl Magnetization Curve for the Half-Filled
Hubbard Model},\ \PTP{42}{1969}{1098}.
\smallskip
\item{10.} K. Penc and F. Woynarovich :\ {\sl Novel magnetic properties
of the Hubbard chain with an attractive interaction},\
\ZPB{85}{1991}{269}.
\smallskip
\item{11.} M. Takahashi :\ {\sl One-Dimensional Hubbard Model at Finite
Temperature},\ \PTP{47}{1972}{69}.
\smallskip
\item{12.} M. Takahashi :\ {\sl Low Temperature Specific Heat of
One-Dimensional Hubbard Model},\ \PTP{52}{1974}{103}.
\smallskip
\item{13.} F.H.L. E\char'31ler, V.E. Korepin and K. Schoutens :\
{\sl Completeness of the SO(4) extended Bethe ansatz for the
one-dimensional Hubbard Model},\ \NPB{384}{1992}{431}.
\smallskip
\item{14.} B.S. Shastry :\ {\sl Decorated Star-Triangle Relations and Exact
Integrability of the One-Dimensional Hubbard Model},\ \JSP{50}{1988}{57}.
\smallskip
\item{15.} T. Koma :\ {\sl An Extension of the Thermal Bethe Ansatz},\
\PTP{83}{1990}{655}.
\smallskip
\item{16.} H. Tsunetsugu :\ {\sl Temperature Dependence of Spin
Correlation Length of Half-Filled One-Dimensional Hubbard Model},\
\JPSJ{60}{1991}{1460}.
\smallskip
\item{17.} A.M. Finkel'shte{\v\i}n :\ \hskip -2pt{\sl Correlation
functions in the one-dimensional Hubbard model},\
\JETPL{25}{1977}{73}.
\smallskip
\item{18.} H.J. Schulz :\ {\sl Correlated Fermions in One Dimension},\
\IJMPB{5}{1991}{57}.
\smallskip
\item{19.} J.M.P. Carmelo and A.A. Ovchinnikov :\ {\sl Generalization
of the Landau Liquid Concept: Example of the Luttinger Liquid}, \
\JPCM{3}{1991}{757}.
\smallskip
\item{20.} F. Woynarovich and H.P. Eckle :\ {\sl Finite-size corrections
for the low lying states of a half-filled Hubbard chain},\
\JPA{20}{1987}{L443}.
\smallskip
\item{21.} N.M. Bogoliubov and V.E. Korepin :\ {\sl The role of
quasi-one-dimensional structures in high-$T_c$ superconductivity},\
\IJMPB{3}{1989}{427}.
\smallskip
\item{22.} F. Woynarovich :\ {\sl Finite-size effects in a
non-half-filled Hubbard chain},\ \JPA{22}{1989}{4243}.
\smallskip
\item{23.} H. Frahm and V.E. Korepin :\ {\sl Correlation functions of
the one-dimensional Hubbard model in a magnetic field},\
\PRB{43}{1991}{5653}.
\smallskip
\item{24.} B.S. Shastry and B. Sutherland :\ {\sl Twisted Boundary
Conditions and Effective Mass in Heisenberg-Ising and Hubbard Rings},\
\PRL{65}{1990}{243}.
\smallskip
\item{25.} M. Fowler and N. Yu :\ {\sl Persistent Current of a Hubbard
Ring Threaded with a Magnetic Flux}, \ \PRB{45}{1992}{11795} .

\vfill\newpage

\section{\sc II. Reprints on the Supersymmetric $t$-$J$ Model}
\item{1.} B. Sutherland :\ {\sl Model for a multicomponent quantum
system},\ \PRB{12}{1975}{3795}.
\smallskip
\item{2.} P. Schlottmann :\ {\sl Integrable narrow-band model with
possible relevance to heavy-fermion systems},\ \PRB{36}{1987}{5177}.
\smallskip
\item{3.} P.A. Bares and G. Blatter :\ {\sl Supersymmetric $t$-$J$
Model in One Dimension: Separation of Spin and Charge},\
\PRL{64}{1990}{2567}.
\smallskip
\item{4.} S. Sarkar :\ {\sl The supersymmetric $t$-$J$ model in one
dimension},\ \JPA{24}{1991}{1137}.
\smallskip
\item{5.} N. Kawakami and S.-K. Yang :\ {\sl Luttinger liquid
properties of highly correlated electron systems in one dimension},\
\JPC{3}{1991}{5983}.
\smallskip
\item{6.} A. F\"orster and M. Karowski :\ {\sl Algebraic properties of
the Bethe ansatz for an spl(2,1)-supersymmetric $t$-$J$ model},\
\NPB{396}{1993}{611}.

\vfill\newpage

\section{\sc III. Reprints on other Models of Strongly Correlated Electrons}
\subsection{\sc A. Models with Local Interaction}
\vskip .5cm
\item{1.} C.N. Yang : \ {\sl Some Exact Results for the Many-Body
Problem in One Dimension with Repulsive Delta-Function Interaction},\
\PRL{19}{1967}{1312}.
\smallskip
\item{2.} M. Takahashi : \ {\sl One-Dimensional Electron Gas with
Delta-Function Interaction at Finite Temperature} ,\
\PTP{46}{1971}{1388}.
\smallskip
\item{3.} D.C. Mattis, E.H. Lieb :\ {\sl Exact Solution of a
Many-Fermion System and Its Associated Boson Field},\
\JMP{6}{1965}{304}.
\item{4.} F.D.M. Haldane :\ {\sl `Luttinger liquid theory' of
one-dimensional quantum fluids: I. Properties of the Luttinger model
and their extension to the general 1D interacting spinless Fermi
gas},\ \JPC{14}{1981}{2585}.
\smallskip
\item{5.} F.D.M. Haldane :\ {\sl Demonstration of the ``Luttinger
Liquid'' character of Bethe-Ansatz-soluble models of 1-D Quantum
Fluids},\ \PLA{81}{1981}{153}.
\smallskip
\item{6.} R.Z. Bariev :\ {\sl Integrable spin chain with two- and
three-particle interactions},\ \JPA{24}{1991}{L549}.
\smallskip
\item{7.} R.Z. Bariev, A. Kl\"umper, A. Schadschneider and J. Zittartz
:\ {\sl Exact solution of a one-dimensional model of hole
superconductivity},\ \JPA{26}{1993}{1249}.
\smallskip
\item{8.} F.H.L. E\char'31ler, V.E. Korepin and K. Schoutens :\ {\sl
New Exactly Solvable Model of Strongly Correlated Electrons Motivated
by High $T_c$ Superconductivity},\ \PRL{68}{1992}{2960}.
\smallskip
\subsection{\sc B. Models with Long Range Interaction}
\vskip .5cm
\item{1.} B. Sutherland :\ {\sl Quantum Many-Body Problem in One
Dimension : Ground State},\ \JMP{12}{1971}{246}.
\smallskip
\item{2.} B. Sutherland :\ {\sl Quantum Many-Body Problem in One
Dimension : Thermodynamics},\ \JMP{12}{1971}{251}.
\smallskip
\item{3.} Y. Kuramoto, H. Yokoyama :\ {\sl Exactly Soluble
Supersymmetric $t$-$J$-Type Model with Long-Range Exchange and
Transfer},\ \PRL{67}{1991}{1338}.
\smallskip
\item{4.} N. Kawakami :\ {\sl Asymptotic Bethe ansatz: Application to
the one-dimensional $t$-$J$ model with long-range exchange and
transfer},\ \PRB{45}{1992}{7525}.
\smallskip

\vfill\newpage
\begin{putreferences}
\centertext{\bfs REFERENCES}

\reference{pw}{P.W. Anderson,\ \sci{235}{1987}{1196}.}
\reference{baxter}{R.J. Baxter,\ {\sl Exactly Solved Models in
Statistical Mechanics}, New York : Academic Press, (1982).}
\reference{vladb}{V.E. Korepin, G. Izergin and N.M. Bogoliubov,\ {\it
Quantum Inverse Scattering Method, Correlation Functions and Algebraic
Bethe Ansatz}, Cambridge University Press, 1993}
\reference{fadtak}{L.D. Faddeev, L.A. Takhtajan,\ {\sl Hamiltonian
Methods in the Theory of Solitons},\ Springer-Verlag, (1987).}
\reference{izs}{C. Itzykson, H. Saleur, J.B. Zuber,\ {\sl Conformal
Invariance and Applications to Statistical Mechanics},\ Singapore:
World Scientific, (1986).}
\reference{is}{Yu.A. Izyumov, Yu.N. Skryabin,\ {\sl Statistical
Mechanics of Magnetically Ordered Systems},\ Engl. Transl.; New York:
Plenum, (1990).}
\reference{jimbo}{M. Jimbo,\ {\sl Yang-Baxter Equation in Integrable
Systems},\ Advanced Series in Mathematical Physics v.10, Singapore:
World Scientific, (1990).}
\reference{liebmat}{E.H. Lieb, D.C. Mattis (eds.),\ {\sl Mathematical
Physics in One Dimension},\ New York: Academic Press, (1966).}
\reference{mw}{B.M. McCoy, T.T. Wu,\ {\sl The Two-Dimensional Ising
Model},\ Cambridge, Mass.: Harvard University Press, (1973).}
\reference{shas}{B.S. Shastry, S.S. Jha, V. Singh (eds.),\ {\sl
Exactly Solvable Problems in Condensed Matter and Relativistic field
Theory},\ Lecture Notes in Physics, v.242, Berlin: Springer Verlag,
(1985).}
\reference{aff}{I. Affleck in talk given at the Nato Advanced
Study Institute on {\sl Physics, Geometry and Topology}, Banff, August
1989.}
\reference{eks5}{F.H.L. E\char'31ler, V.E. Korepin, K. Schoutens,\
 \PRL{70}{1993}{73}.}
\reference{Gau}{M.Gaudin,\ La Fonction d'Onde de Bethe,\ \ Masson, 1983.}
\reference{Gau2}{M.Gaudin,\ \PLA{24}{1967}{55}.}
\reference{wiegm}{A.M. Tsvelick, P.B. Wiegmann,\ \AdP{32}{1983}{453}.}
\reference{low}{N. Andrei, K. Furuya, J.H. Lowenstein,\
\RMP{55}{1983}{331}.}
\reference{ya1}{C.N. Yang,\ \PRL{19}{1967}{1312}.}
\end{putreferences}
\bye